\documentclass{article}
\usepackage[utf8]{inputenc}
\usepackage{jheppub}
\usepackage{amsmath}
\usepackage{amssymb}
\usepackage{amsfonts}
\usepackage{amsthm}
\usepackage{physics}
\usepackage{appendix}
\usepackage{float}
\usepackage{cancel}
\makeatletter
\def\moverlay{\mathpalette\mov@rlay}
\def\mov@rlay#1#2{\leavevmode\vtop{%
   \baselineskip\z@skip \lineskiplimit-\maxdimen
   \ialign{\hfil$\m@th#1##$\hfil\cr#2\crcr}}}
\newcommand{\charfusion}[3][\mathord]{
    #1{\ifx#1\mathop\vphantom{#2}\fi
        \mathpalette\mov@rlay{#2\cr#3}
      }
    \ifx#1\mathop\expandafter\displaylimits\fi}
\makeatother

\newcommand{\cupdot}{\charfusion[\mathbin]{\cup}{\cdot}}

\def\b1{\bar 1}

\def\g{\mathfrak{g}}
\def\Lc{\mathcal{L}}
\def\Hc{\mathcal{H}}
\def\cG{\mathcal{G}}
\def\dep{\mathfrak{d}} 
\title{Gaudin Models and Multipoint Conformal Blocks: General Theory}

\subheader{\normalsize
        \vspace*{-3.7em}
        \begin{flushright}
	DESY 21-052\\
        SAGEX 21-08-E\\
        ZMP-HH/21-8
        \end{flushright}
}

\author{Ilija Buri\'c$^1$,}
\author{Sylvain Lacroix$^2$,} 
\author{Jeremy A. Mann$^1$,} 
\author{Lorenzo Quintavalle$^1$,} 
\author{Volker Schomerus$^{1,2}$}
\affiliation{$^1$ DESY Theory Group, DESY Hamburg, Notkestrasse 85, D-22603 Hamburg,}
\affiliation{$^2$ II. Institut f\"ur Theoretische Physik, Universit\"at Hamburg, Luruper Chaussee 149, D-22761 Hamburg}
\affiliation{Zentrum f\"ur Mathematische Physik, Universit\"at Hamburg, Bundesstrasse 55, D-20146 Hamburg }

\emailAdd{ilija.buric@desy.de}
\emailAdd{sylvain.lacroix@desy.de}
\emailAdd{jeremy.mann@desy.de}
\emailAdd{lorenzo.quintavalle@desy.de}
\emailAdd{volker.schomerus@desy.de}

\date{April 2021}

\abstract{The construction of conformal blocks for the analysis of multipoint 
correlation functions with $N > 4$ local field insertions is an important open 
problem in higher dimensional conformal field theory. This is the first in a 
series of papers in which we address this challenge, following and extending 
our short announcement in \cite{Buric:2020dyz}.  According to Dolan and Osborn, 
conformal blocks can be determined from the set of differential eigenvalue 
equations that they satisfy. We construct a complete set of commuting differential 
operators that characterize multipoint conformal blocks for any number $N$ of 
points in any dimension and for any choice of OPE channel through the relation
with Gaudin integrable models we uncovered in \cite{Buric:2020dyz}. For 5-point 
conformal blocks, there exist five such operators which are worked out smoothly 
in the dimension $d$.}

\begin{document}
\addtolength{\baselineskip}{2mm}
\maketitle

\section{Introduction and Summary of Results}
\label{sec:sum}

In the last few decades conformal field theories have steadily gained importance, first 
in $d=2$ and then for $d > 2$. They provide a powerful source of new paradigms that advance 
our understanding of quantum field theory deep in the quantum regime which is inaccessible 
to perturbation theory. Through the AdS/CFT correspondence conformal field theories have
even begun to teach us about some of the deepest mysteries of quantum gravity. 

In a conformal field theory, Poincare symmetry is enhanced to conformal symmetry. This symmetry
enhancement turns out to be highly constraining, even in $d>2$, where the conformal group is 
finite dimensional and only contains $d+1$ generators outside of the Poincare group. These few 
additional generators provide 
enormous control over the operator product expansion (OPE) of local fields and in fact 
fix these products up to some numerical OPE coefficients. The latter determine all higher 
correlations, at least in principle. In practice, however, higher correlations with more 
than $N=3$ field insertions possess an intricate dependence on the insertion points
of the local fields and their simplicity only becomes visible after expanding correlators
in a basis of conformal blocks. The latter are very much like plane waves in ordinary 
Fourier analysis, i.e.\ conformal blocks (or rather the closely related conformal 
partial waves) are to conformal symmetry what plane waves are to the translation 
group. Once the basis of conformal blocks is known, one can expand the correlation 
function. Conformal symmetry then implies that the coefficients are simply products 
of the coefficients that appear in the OPE. 

All this is of course well known since the early days of conformal field theory, see 
\cite{Ferrara:1973vz,Ferrara:1973yt,Mack:1976pa}. But the pioneers of conformal field 
theory were only able to write down integral formulas for conformal blocks 
\cite{Ferrara:1972uq,Dobrev:1977qv}. These so-called shadow integrals are very 
much like Feynman integrals. In particular, it is a considerable challenge to 
relate shadow integrals to some known special functions, to read off their properties 
and to find relations between them, at least for $d > 2$ where the integrations cannot 
be performed directly. There has been little progress on this problem until 
Dolan and Osborn shifted attention from the integrals to the differential equations that these integrals satisfy \cite{Dolan:2000ut,Dolan:2003hv}. Once they had constructed 
their so-called Casimir differential operators they were able to extract a wealth of 
information on $4$-point blocks, see also \cite{Dolan:2011dv,Hogervorst:2013sma,Penedones:2015aga}. 
This work has been the decisive input for the modern conformal 
bootstrap program (see \cite{Poland:2018epd} for a review and many references to the original 
literature) and its spectacular applications to the $d=3$ Ising model, in particular 
\cite{ElShowk:2012ht,El-Showk:2014dwa,Kos:2016ysd,Simmons-Duffin:2016wlq}. More 
recently, it was observed that the Casimir differential 
operators studied by Dolan and Osborn could be identified with the Hamiltonians of 
an integrable 2-particle quantum mechanics system of Calogero-Sutherland type 
\cite{Isachenkov:2016gim}. The observation was later explained in the context of 
harmonic analysis for the conformal group \cite{Schomerus:2016epl,Schomerus:2017eny,
Buric:2019rms,Buric:2020buk}. Calogero-Sutherland Hamiltonians and their eigenfunctions 
have actually been studied in mathematics for several decades where it has become 
instrumental in developing the modern theory of multivariate hypergeometric functions. After 
the relation with conformal blocks was noticed it became clear that most of the 
results on $4$-point blocks had been known long before the work in conformal field 
theory, mostly from the early work on Calogero-Sutherland models by Heckman and Opdam 
\cite{Heckman-Opdam}, see \cite{Isachenkov:2017qgn}. 

In principle it is possible to probe a conformal field theory by studying the entire 
set of 4-point functions for the infinite number of (spinning) conformal primary fields. 
And even though the theory of spinning conformal blocks is reasonably well developed
by now, see e.g. \cite{Costa:2011dw,Costa:2011mg,Echeverri:2016dun,Karateev:2017jgd}, 
working with an infinite number of correlation functions seems impossible at least in 
the absence of any additional algebraic structure. In practice, it may appear more 
promising to study higher correlations of a small set of fundamental fields of the 
theory. The $8$-point function of a fundamental scalar field, for example, already contains as much dynamical information as an infinite set of spinning $4$-point functions. 
Given that it took more than 30 years to develop a theory of $4$-point blocks, it may seem 
like a daunting task to actually extend this theory to multi-point blocks, though there 
has been significant activity in this direction lately, see e.g. \cite{Rosenhaus:2018zqn,
Parikh:2019ygo,Fortin:2019dnq,Parikh:2019dvm,Fortin:2019zkm,Irges:2020lgp,Fortin:2020yjz,
Pal:2020dqf,Fortin:2020bfq,Hoback:2020pgj,Goncalves:2019znr,Anous:2020vtw,Fortin:2020zxw,
Poland:2021xjs}. The embedding of $4$-point blocks into harmonic analysis of the conformal 
group and the profound relations with integrable quantum mechanical models may be viewed 
as a strong hint towards the appropriate mathematical tools for the study of multi-point 
blocks. And in fact, as we have announced in \cite{Buric:2020dyz}, the relation to 
integrable models is not restricted to the case of $4$-point blocks. As we have outlined, 
multi-point blocks for any number $N$ of local fields can be characterized as joint 
eigenfunctions of a complete set of commuting differential operators. The latter were 
argued to arise as Hamiltonians of Gaudin integrable models. 
\medskip 

Our goal in this paper and its sequels is to substantiate our claims and to develop 
them into a full theory of multi-point blocks. In this paper we perform the first 
step of this program and explain in detail the relation between multi-point conformal 
blocks and Gaudin integrable systems. We will actually go further than \cite{Buric:2020dyz} which only sketched the construction of differential operators
for the so-called comb channel conformal blocks \cite{Rosenhaus:2018zqn}. The results 
described in this work apply to any channel, i.e.\ they provide a complete set of 
commuting differential operators for whatever channel one is interested in, including 
e.g. the snowflake channel that has also received some attention lately, 
\cite{Fortin:2020yjz,Fortin:2020bfq}. As we shall explain below, 
in any given channel, the differential operators split into two groups: the Dolan-Osborn-like Casimir differential operators that measure the weight and spin of intermediate 
fields, and the novel vertex differential operators that measure a choice of tensor 
structure in the operator product of intermediate (spinning) fields. In a second 
paper of this series we will address such internal vertices for the comb channel 
of 3- and 4-dimensional conformal field theories. These vertices give rise to a 
single vertex differential operator, see \cite{Buric:2020dyz} for an example, that 
we will identify as the Hamiltonian of 
an elliptic Calogero-Moser system for some complex crystallographic reflection group, first obtained 
in \cite{etingof2021elliptic}. Extensions to vertices with more than one degree of freedom, which are relevant e.g. for the snowflake channel in $4D$, as well as a theory of 
solutions for single and multi-variable vertices will be addressed in the future.  
\medskip 

The main results of this work are the following. We construct the Casimir and 
vertex differential operators that are simultaneously diagonalized by conformal blocks. 
Commutativity of this set of operators is established by realizing it as a limit of the 
Gaudin integrable model. The connection to the Gaudin model seems to be necessary for the 
full proof, although some parts of the proof can be established by elementary arguments. 
Further, we provide evidence that independent Casimir and vertex operators are equal in 
number to independent cross ratios and can thus be used to completely characterize conformal blocks. 
This is done by exhibiting a number of relations between the operators, leaving us with 
an explicit commuting system with no apparent further dependencies. Finally, we construct 
the five commuting Casimir and vertex operators for $5$-point functions explicitly as
differential operators in the five cross ratios. 

For the remainder of this introduction we will give a more precise summary of the main 
results and place them in the context of conformal field theory. 
\medskip 

\noindent 
\textbf{Summary of Results} 
In this work we consider correlation functions of $N$ local scalar primary fields 
$\phi_i$ with conformal weights denoted by $\Delta_i$,  
\begin{equation}
\cG_N(x_i,\Delta_i) := \langle 0| \phi_1(x_1) \cdots \phi_N(x_N)| 0 \rangle \ .      
\end{equation}
The fields are inserted at points $x_i, i=1, \dots, N$, of $d$-dimensional 
Euclidean space $\mathbb{R}^d$. Scalar primary fields are characterized by their
commutation relations with the generators of the conformal algebra which take the 
form
\begin{equation} 
[T_\alpha,\phi_i(x_i)] = {\mathcal{T}}_{\alpha}^{(i)} \phi_i(x_i),
\end{equation} 
where $\cal{T}_\alpha$ are the usual first order differential operators 
 and $\alpha$ runs through the set of conformal generators. 

\begin{figure}[thb]
\centering
\includegraphics{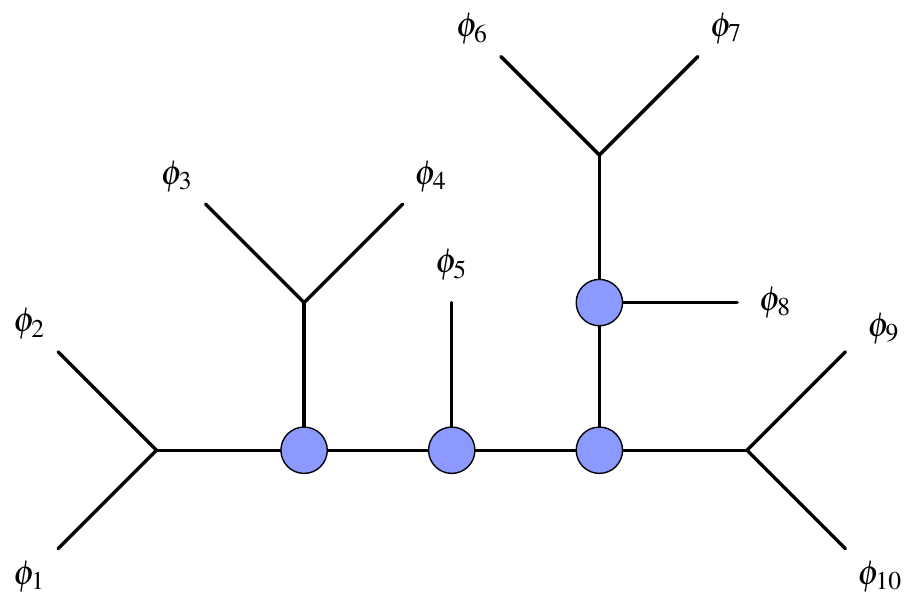}
\caption{Choice of an OPE channel for a 10-point function. Vertices highlighted in blue are associated with non-trivial tensor structures.}
\label{fig:10pointOPE}
\end{figure}

\noindent 
\textit{Prerequisites and notation.} 
Conformal correlation functions are symmetric with respect to the exchange of any two 
fields. On the other hand, the way we compute them is not. In fact, $N$-point correlation 
functions are evaluated by a repeated use of operator product expansions (OPEs) and the 
precise sequence of performing these expansions determines what is known as an OPE channel. 
These channels can be represented as (plane) tree diagrams $\mathcal{C}^N_\textit{OPE}$ 
with enumerated leaves. Figure~\ref{fig:10pointOPE} shows one such example for the 10-point function. Given 
a fixed enumeration of the external scalar fields, the number of OPE channels is 
$(2N-5)!!$. Each of these channels is associated with a system of conformal blocks that 
provide a Fourier-like expansion of the correlation function. After stripping off some 
appropriate prefactors $\Omega(x_i|\Delta_i)$, the conformal blocks depend only on 
conformally invariant cross ratios. It is well known that one can form 
\begin{equation}
n_\textit{cr} (N,d) = \left\{ \begin{array}{ll} \frac12 N (N-3) \quad & N \leq d + 2 \\[3mm] 
Nd - \frac12(d+2)(d+1) \quad &  N > d + 2 \end{array} \right.  
\end{equation} 
of such independent cross ratios from a set of $N$ points $x_i \in \mathbb{R}^d$. Note 
that this number is independent of the OPE channel. The conformal blocks we want to 
expand our correlation functions in must depend on the same number of labels. For example, 
in the case of a 4-point function in $d>1$ one has two cross ratios. The associated 
blocks are famously labeled by the conformal weight $\Delta$ and the spin $l$ of the 
field that is exchanged in the intermediate channel. Since the latter transforms in a 
symmetric traceless tensor representation of the rotation subgroup, a single number 
$l$ is sufficient to characterize the spin. 

When dealing with higher correlation functions, it is easy to see that the quantum numbers of fields exchanged in the intermediate channels are not sufficient. The precise number of such intermediate field labels does depend on the channel topology,
at least for $N > 5$, but it is always strictly smaller than the number $n_\textit{cr}$ 
of cross ratios. As an example let us consider $N=5$. 
\begin{figure}[thb]
\centering
\includegraphics{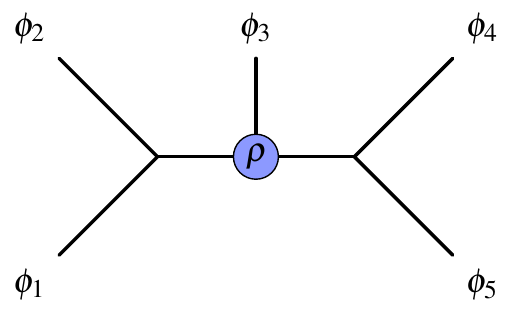}
\caption{Choice of OPE diagram for 5-point correlator.}
\label{fig:FivePointsOPE}
\end{figure}
In this case there exist $15$ 
OPE channels but they all possess the same topology. In Figure~\ref{fig:FivePointsOPE} we have displayed 
one of the fifteen OPE channels. All other 14 channels are related to this one by 
a permutation of the leaves, modulo symmetries of the bare OPE diagram that has 
been stripped of its leaves. As one can readily see, the OPE evaluation involves two 
intermediate fields. Since these appear in the operator product of scalar fields, they 
are symmetric traceless tensors and hence characterized by two quantum numbers each. 
The intermediate fields thus carry four quantum numbers in total. We think of these 
as being attached to the (internal) links of the OPE diagram. But in order to fully 
characterize the relevant blocks we need five quantum numbers and so we are short by 
one. As we shall show below, this remaining quantum number is attached to the central 
vertex of the 5-point OPE diagram. Note that two legs of this central vertex are 
associated with a symmetric traceless tensor while only one is scalar. For $d\geq 3$, 
such 3-point functions are not determined by conformal symmetry. They require the 
choice of a so-called tensor structure. For a very particular basis in the space of 
tensor structures it becomes possible to assign a fifth quantum number to the 
choice of tensor structure, one that can be measured simultaneously with the 
four independent eigenvalues of the Casimir operators. Measuring a complete 
set of quantum numbers simultaneously through a sufficiently large set of 
commuting differential operators is the main goal of this work. We want to 
do so in any $d$, for any number $N$ of external points and for all OPE channels. 
Quantum numbers are measured by acting with differential operators in the cross 
ratios. The latter divide into two families. First, differential operators that measure 
the quantum numbers of intermediate fields are associated with the 
links of the OPE diagram and are referred to as \textit{Casimir differential operators},
since they are straightforward generalizations of the Casimir differential operators 
constructed for $N=4$ by Dolan and Osborn. Second, differential operators that measure choices of tensor structure, the first example of which was introduced recently in \cite{Buric:2020dyz}, are referred to as \textit{vertex 
differential operators}. Let us note that for scalar blocks, the choice of tensor structures and hence the vertex differential operators are relevant as soon as multiple non-scalar exchanges are involved. These types of blocks have only been considered very 
recently in \cite{Poland:2021xjs} and \cite[Appendix E]{Goncalves:2019znr} for five-point blocks, and in a 
certain limit in \cite{Vieira:2020xfx} for five or six scalar legs. 

Before we can describe our results on both types of differential operators we need 
to set up some notation. Given an OPE channel we enumerate internal lines by Latin 
indices $r = 1 , \dots, N-3$ and vertices by Greek indices $\rho = 1, \dots, N-2$. 
There are no rules on how to enumerate these two sets of objects. OPE diagrams are 
(plane) trees and hence by cutting any internal line with label $r$ we separate the 
diagram into two disconnected pieces. Hence, $r$ is associated with a partition of 
the external fields into two disjoint sets, 
\begin{equation}
\underline{N} = \{ 1, \dots, N\} = I_{r,1} \cupdot I_{r,2}\ .       
\end{equation} 
Similarly, any vertex $\rho$ gives rise to a partition of $\underline N$ into three 
disjoint sets 
\begin{equation} \label{eq:NIrho}
\underline{N} = I_{\rho,1} \cupdot I_{\rho,2} \cupdot I_{\rho,3}\ . 
\end{equation} 
Given any subset $I \subset \underline{N}$ we can define the following set of 
first order differential operators in the insertion points $x_i$, 
\begin{equation} \label{eq:TPI}
\mathcal{T}^{(I)}_\alpha  = \sum_{i\in I} \mathcal{T}^{(i)}_\alpha\ .       
\end{equation}
Let us note that for two disjoint sets $I_1, I_2 \subset \underline N$ we have
\begin{equation}  \label{eq:commdisjoint} 
\mathcal{T}^{(I_1 \cupdot I_2)}_\alpha = \mathcal{T}^{(I_1)}_\alpha + 
\mathcal{T}^{(I_2)}_\alpha \ , 
\quad [ \, \mathcal{T}^{(I_1)}_\alpha \, , \, \mathcal{T}^{(I_2)}_\beta \, ] = 0 \ .  
\end{equation} 

\noindent 
\textit{Casimir differential operators.} 
With this notation it is very easy to construct the differential operators that measure 
the quantum numbers of the intermediate fields,  
\begin{equation} \label{eq:Casdiffop} 
\mathcal{C}\textit{\!as\,}^p_r = \mathcal{D}^{p}_{r,1}  =  
\kappa_p^{\alpha_1,\dots,\alpha_{p}}\left[\mathcal{T}^{(I_{r,1})}_{\alpha_1}
    \cdots \mathcal{T}^{(I_{r,1})}_{\alpha_p}\right]_{|\cG} = \mathcal{D}^p_{r,2}\ .       
\end{equation}
Here $\kappa_p$ denotes \textit{symmetric} conformally invariant tensors of order $p$ and 
the superscript $p$ runs through 
\begin{equation} 
p = 2, 4, \dots \left\{ \begin{array}{ll} d+1 = 2 r_d  \quad & \textit{for $d$ odd} \\[2mm]  
d = 2 r_d -2 \quad & \textit{for $d$ even} \end{array} \right. 
\end{equation} 
The number $r_d = [(d+2)/2]$ denotes the rank of the conformal Lie algebra. 
In even dimensions $d$, the symmetric invariant tensor $\kappa_p$ of order $p = 2r_d = d+2$ 
actually possesses a square root of order $p = r_d$ that also commutes with all generators of 
the conformal algebra. This so-called \textit{Pfaffian} differential operator has the same form 
as in \eqref{eq:Casdiffop}, but with a symmetric invariant tensor $\kappa_p$ of order $p=d/2+1$, 
\begin{equation} \label{eq:PfCasimir} 
\mathcal{P}\textit{\!f\,}_{r} = \mathcal{D}^{d/2+1}_{r,1} = 
    \kappa_{d/2+1}^{\alpha_1,\dots,\alpha_{d/2+1}} \left[\mathcal{T}^{(I_{r,1})}_{\alpha_1}
    \cdots \mathcal{T}^{(I_{r,1})}_{\alpha_{d/2+1}}\right]_{|\cG}  = - (-1)^{d/2} 
    \mathcal{D}^{d/2+1}_{r,2}\ . 
\end{equation} 
When $d=4k+2$, the symmetric invariants of order $p = d/2+1$ are twofold degenerate and we 
should use two different symbols for these two invariants of order $d/2+1$. In order not 
to clutter notation too much, we decided to ignore this distinction. In other words, we 
will consider $\kappa_{d/2+1}$ as a pair of symmetric invariants when $d=4k+2$.  

In our formulas for the differential operators we have placed a subscript $|\cG$ to stress 
that they are defined as operators acting on correlations functions, i.e. on functions $\cG$ 
that satisfy the conformal Ward identities
\begin{equation}\label{eq:ward}
\mathcal{T}^{(\underline{N})}_\alpha \mathcal{G}_N(x_i,\Delta_i) = 0 \  .      
\end{equation}
In our construction of the differential operators we have favored the set $I_{r,1}$ over 
$I_{r,2}$. But from the conformal Ward identities we can conclude that
$$ \mathcal{T}^{(I_{r,1})}_\alpha \cG_N(x_i,\Delta_i) =  - \mathcal{T}^{(I_{r,2})}_\alpha 
\cG_N(x_i,\Delta_i) \ . $$ 
Though some caution is needed when we apply this relation to the evaluation of the 
Casimir differential operator, see Subsection~\ref{sec:red} for details, it is not difficult to 
see that all differential operators of even order come out the same if we pick $I_{r,2}$
rather than $I_{r,1}$. There is only one family for which the set matters, namely for the 
Pfaffian operators when $d$ is a multiple of four. In that case the operator flips sign 
when we change the set. Of course, overall factors are a matter of convention and hence 
of no concern. Therefore, we shall drop the reference to the set we use in the 
construction of Casimir differential operators, writing $\mathcal{D}^p_{r}$ instead 
of $\mathcal{D}^p_{r,1}$. 

An important point to note is that the Casimir differential operators need not be 
independent. To illustrate this, consider the case $N=4$ for $d>2$. Since all external 
fields are assumed to be scalar, the single intermediate field is a symmetric traceless 
tensor and is hence characterized by two numbers only, its weight $\Delta$ and spin $l$. 
These can be measured by the Casimir differential operators  $\mathcal{D}^2$ and 
$\mathcal{D}^4$. But starting from $d =4$, the conformal algebra possesses Casimir 
elements of higher order which are independent in general, but become dependent on
the lower order ones when evaluated on symmetric traceless tensors. More generally, 
the number of independent Casimir differential operators at a given internal line 
$r$ is given by 
\begin{equation} \label{eq:depth} 
\dep_r(\mathcal{C}^N_\textit{OPE},d) = \dep(I_{r,1},d), \quad \textit{where} 
\quad \dep(I,d) = \textit{min}(|I|,N-|I|,r_d),
\end{equation} 
and $|I|$ denotes the order of the set $I$.\footnote{See the appendix B, where we collect 
some elements of $SO(1,d+1)$ representation theory.} We shall refer to the number $\dep(I,d)$ 
as the \textit{depth} of the index set $I$ and to $\dep_r$ as the depth of the link $r$. 
Note that $\dep_r = \dep(I_{r,1},d) = \dep(I_{r,2},d)$ is independent of which of the 
two index sets we choose to compute it with. By summing the depths of all internal links, we 
can determine the total number of independent Casimir differential operators to be 
\begin{equation}
n_{\textit{cdo}}(\mathcal{C}^N_\textit{OPE},d) = \sum_{r=1}^{N-3} 
\dep_r(\mathcal{C}^N_\textit{OPE},d)\ . 
\end{equation}
Let us note that the total number of Casimir differential operators does depend
on the topology of the OPE channel, not just on the number $N$ of points. 
In the case of $N=6$ and $d\geq 4$, for example, there are $n_\textit{cdo}(\mathcal{C}^{N=6}_
\textit{comb},d) = 7$ Casimir differential operators in the comb channel, while the 
snowflake channel admits only $n_\textit{cdo}(\mathcal{C}^{N=6}_\textit{snowflake},d) = 
6$ of such operators. 
\medskip 

\noindent 
\textit{Vertex differential operators.} What we have described so far is nothing new, and can be established by elementary means. But as we have explained, 
starting from $N = 5$ the Casimir differential operators do not suffice to resolve 
all quantum numbers of the conformal blocks, i.e.\ $n_{\textit{cdo}}$ is strictly 
smaller that $n_\textit{cr}$ for all OPE channels. Our main task is to construct 
additional differential operators that can measure the choice of tensor structures
at the vertices independently of the weights and spins of the intermediate fields, 
i.e.\ we need to find a complete set of vertex differential operators 
that commute with the Casimir differential operators and among themselves. In 
this work we describe how to accomplish this task, for any number $N$ of external 
scalar fields and any OPE topology. One central claim is that these vertex 
differential operators take the form 
\begin{equation}
    \mathcal{D}^{p,\nu}_{\rho,12}  =   
    \kappa_p^{\alpha_1,\dots,\alpha_{\nu},\alpha_{\nu+1}, \dots, \alpha_p}
    \left[\mathcal{T}^{(I_{\rho,1})}_{\alpha_1}
    \cdots \mathcal{T}^{(I_{\rho,1})}_{\alpha_\nu}
    \mathcal{T}^{(I_{\rho,2})}_{\alpha_{\nu+1}}
    \cdots \mathcal{T}^{(I_{\rho,2})}_{\alpha_p}\right]_{|\cG} \label{eq:vdo} 
\end{equation}
where $\nu = 1, \dots, p-1$ and $p=2, 4, \dots, 2r_d = d+1$ when $d$ is odd. 
For even $d$, we let $p$ run through even integers until we reach $d$ and add
a set of Pfaffian vertex operators $\mathcal{P}\textit{f\;}^\nu_{\rho,12}, \nu = 1,2,
\dots d/2$ which are constructed with a symmetric invariant tensor $\kappa_p$ of order 
$p=d/2 + 1$. Let us note that the definition of all these vertex differential 
operators also makes sense for $\nu=0$ and $\nu = p$. The corresponding objects 
coincide with Casimir differential operators for the links that enter the first 
and second leg of the vertex. This is why we have excluded them from our list. The 
remaining operators still allow us to reconstruct the Casimir operators for the link 
that enters the third leg. Therefore, there is one linear relation for each value that $p$ 
can assume, i.e. we have $r_d$ linear relations in total. One may use these 
relations to eliminate e.g. the operator with $\nu = p/2$.  

Let us note that the definition of the vertex operators $\mathcal{D}^{p,\nu}_{\rho,ij}$ 
depends on the choice of labeling of the subsets $I_{\rho,j}$ forming the partition 
$\underline{N} = I_{\rho,1} \cupdot I_{\rho,2} \cupdot I_{\rho,3}$ associated with 
the vertex $\rho$, which is arbitrary. However, the algebra generated by the vertex 
operators $\mathcal{D}^{p,\nu}_\rho$ is in fact independent of this choice: more 
precisely, the vertex operators constructed from another choice of labeling of 
the $I_{\rho,j}$'s are linear combinations of the operators $\mathcal{D}^{p,
\nu}_{\rho,12}$, modulo the use of the conformal Ward identities \eqref{eq:ward}, 
see Section \ref{sec:3pt}. 

The number of vertex differential operators at a given vertex is now easy to count. 
Taking into account that one additional linear relation among the operators 
listed in eq.\ \eqref{eq:vdo}, one finds 
\begin{equation} \label{eq:novertfull} 
n_v(d) = \left\{ \begin{array}{ll} \frac14(d^2-4) \quad & \textit{for d even}\\[2mm]
\frac14(d^2-1) \quad & \textit{for d odd} \end{array}\right.  
\end{equation} 
The first key result of this work is that these vertex differential operators
commute among themselves and with the Casimir differential operators. Commutation 
between Casimir and vertex differential operators is obvious. Similarly, it is 
easy to show that two vertex operators commute if they are associated with 
different vertices $\rho \neq \rho'$. The deepest part of our claim concerns the 
fact that also vertex operators associated with the same vertex commute. It does 
not seem straightforward to prove this statement by elementary manipulations. Below 
we shall use an indirect strategy in which we identify these vertex differential 
operators with Hamiltonians of some Gaudin integrable system defined on a 
3-punctured sphere. For the latter, commutativity has already been established. 

\begin{figure}[thb]
\centering
\includegraphics{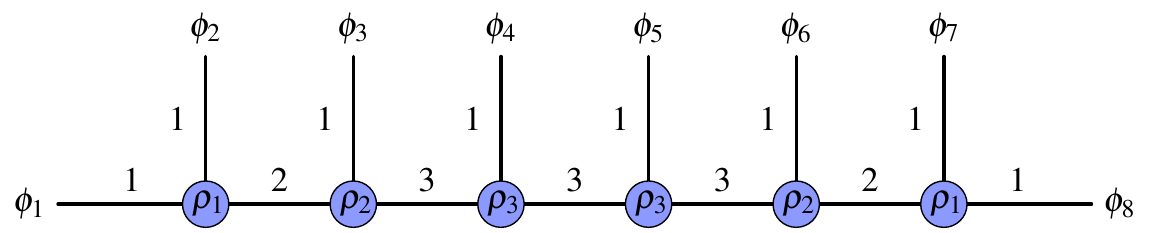}
\caption{OPE diagram in the comb channel for a scalar eight-point function in $d=4$. The edge labels correspond to the depth $\dep$ of the associated links. When taking an OPE with a scalar field, the depth always increases by one until the maximal allowed depth $\dep=r_d$ is reached.}
\label{fig:8pointcomb}
\end{figure}

Of course the vertex differential operators we listed may not all be independent, 
as for the Casimir operators, see discussion above. In order to 
count the number of independent vertex differential operators, we shall employ the 
depth function $\dep(I,d)$ we introduced in eq.\ \eqref{eq:depth}. For a given 
vertex $\rho$ inside an OPE channel $\mathcal{C}^N_\textit{OPE}$, the number of independent vertex differential operators is expected to be equal to the degrees of freedom associated to this vertex
\begin{equation} \label{eq:novertrest} 
n_{\textit{vdo},\rho}(\mathcal{C}^N_\textit{OPE},d) = 
n_\textit{cr}(\sum_{i=1}^3 \dep_{\rho,i},d) - 
\sum_{i=1}^3 \dep_{\rho,i} (\dep_{\rho,i}-1)  \leq n_v(d)
\end{equation}
where $\dep_{\rho,i} = \dep(I_{\rho,i},d)$ with $i =1,2,3$. The 
inequality is saturated for vertices $\rho$ with $\dep_{\rho,i} = r_d$. For 
the special vertices that can appear in the comb channel and in which one of the 
legs is scalar, the formula becomes 
$$ n_{\textit{vdo},\rho_m} = m-1 \  , \quad n_{\textit{vdo},\rho_{r_d}} = 
r_d - 1 - \delta_{d,\textit{even}}
$$ 
for $m = 1, \dots, r_d-1$. Here $\rho_m$ is a vertex with $\dep_{\rho_m,1} = m, 
\dep_{\rho_m, 2} = 1$ and $\dep_{\rho_m,3} = m+1$, see Figure~\ref{fig:8pointcomb}, and $\rho_d$
is the maximal comb channel vertex with $\dep_{\rho_{r_d},1} = r_d = 
\dep_{\rho_{r_d},3}$. The total number $n_\textit{vdo}(\mathcal{C}^N_\textit{OPE},d)$ 
of vertex differential operators is obtained by summing over all $N-2$ vertices, i.e. 
$$n_\textit{vdo}(\mathcal{C}^N_\textit{OPE},d) = \sum_{\rho = 1}^{N-2} 
n_{\textit{vdo},\rho}(\mathcal{C}^N_\textit{OPE},d)\ . $$ 
At least for the comb channel, it is easy to verify that the number of independent 
Casimir and vertex differential operators coincides with the number of cross ratios, 
$$  n_{\textit{cdo}}(\mathcal{C}^N_\textit{OPE},d) +
n_\textit{vdo}(\mathcal{C}^N_\textit{OPE},d) = n_\textit{cr}(N,d) \ . $$
The formula holds of course for all OPE channels. Below we shall exhibit the 
relations among vertex differential operators that are responsible for the 
reduction from the $n_v(d)$ operators in our list \eqref{eq:vdo} (with $\nu = 
p/2$ removed) to the $n_{\textit{vdo},\rho} (\mathcal{C}^N_\textit{OPE},d)$ 
independent vertex differential operators that are needed to characterize 
the vertex $\rho$. This is the second key result of this work. It will allow 
us in particular to determine the precise order of each independent 
vertex differential operator. 

While Gaudin models for the 3-punctured sphere only enter the discussion as 
a convenient tool to construct commuting vertex differential operators at the 
individual vertices, the relation between conformal blocks and Gaudin models 
turns out to reach much further. In fact, is is possible to embed the whole 
set of Casimir and vertex differential operators for arbitrary scalar $N$-point 
functions into Gaudin models on the $N$-punctured sphere. The latter contains 
$N$ additional complex parameters that are not present in correlation 
functions. In the Gaudin integrable model these parameters correspond to 
the poles of the Lax matrix and they enter all Gaudin Hamiltonians. By 
considering different limiting configurations of these parameters is is 
possible to recover the full set of Casimir and vertex differential 
operators for scalar $N$-point functions in all the OPE channels. This 
construction not only embeds our differential operators into a unique 
Gaudin integrable model, but also shows that operators in different channels 
are related by a smooth deformation.
\medskip 

Let us finally outline the content of the following sections. 
Section~\ref{sect:vertexsystem} is mostly devoted to the study of the 
individual vertices. After a brief discussion of commutativity for Casimir 
differential operators and also vertex differential operators assigned to 
different vertices, we shall zoom into the individual vertices for most of Section~\ref{sect:vertexsystem}. In Section~\ref{sec:3pt} 
we construct the vertex differential operators in terms of the commuting 
Hamiltonians of a 3-site Gaudin integrable system. Section~\ref{sec:dependences} 
addresses the relations between these operators for restricted vertices. 
The main purpose of Section~\ref{sec:GaudinOPE} is to embed the whole set of 
Casimir and vertex differential operators for arbitrary scalar $N$-point 
functions into Gaudin models on the $N$-punctured sphere. In 
Section~\ref{section:fivepoints} we discuss one concrete example, namely we 
construct all five differential operators that characterize the blocks of a scalar 
5-point function in any $d \geq 3$. Two of these operators are of 
order two while the other three are of fourth order. The paper 
concludes with a summary, outlook to further results  and a list 
of interesting open problems to be addressed.

\section{The Vertex Integrable System}
\label{sect:vertexsystem}

The aim of this section is to address the key new element in the construction 
of multi-point conformal blocks for $d \geq 3$: the vertices themselves. In the first subsection we shall show that the 
construction of commuting differential operators for scalar $N$-point 
blocks can be reduced by rather elementary arguments to the construction of 
commuting differential operators for 3-point functions of spinning fields. 
Recall that the dependence on spin degrees of freedom can be encoded in 
auxiliary variables from which one is often able to construct non-trivial 
cross ratios,\footnote{Here and throughout the entire text we use the term 
`cross ratio' rather loosely to refer to all conformally invariant 
combinations of the positions and auxiliary variables at each point.} even 
in the case of a 3-point function. Constructing sufficiently
many commuting vertex differential operators that act on such cross ratios of 
the 3-point function requires more powerful technology from integrability which 
we shall turn to in the second subsection. 
There we construct commuting differential operators for vertices 
from the Gaudin integrable model for the 3-punctured sphere. The 
basic construction provides $n_v(d)$ of such commuting operators and 
hence sufficiently many even for the most generic vertices. For 
special vertices, such as those appearing in the comb channel with external scalars, 
there exist linear relations between these operators. These are the 
subject of the third subsection. 

\subsection{Reduction to the vertex systems}
\label{sec:red}

Our goal here is to prove that Casimir and vertex operators constructed around 
different vertices of an OPE diagram commute. More precisely we shall show that 
\begin{equation} \label{eq:commCDO} 
[\, \mathcal{D}^{p}_r\,, \, \mathcal{D}^{q}_{r'} \, ]  = 0 \quad , \quad
[\, \mathcal{D}^p_r\, ,\, \mathcal{D}^{q,\nu}_{\rho,a} \,] = 0 
\end{equation} 
for every pair $(r,p),(r',q)$ of Casimir operators and any choice $(\rho,q,\nu)$ 
of a vertex operator, including the Pfaffian Casimir and vertex operators that 
appear at order $p,q=d/2 + 1$ when the dimension $d$ is even. Note that the 
individual vertex differential operators also depend on the choice $a \in 
\{12,23,13\}$ of a pair of legs. In addition, we shall also establish that 
vertex operators associated with different vertices commute, 
\begin{equation} \label{eq:commVDO1}
[\, \mathcal{D}^{p,\nu}_{\rho,a}\, ,\, \mathcal{D}^{q,\mu}_{\rho',b} \,] = 0 \quad 
\mathit{ for } \ \rho \neq \rho'
\end{equation}
and all triples $(p,\nu,a)$ and $(q,\mu,b)$. This leaves only the commutativity of operators attached to the same vertex which is deferred to the next subsection. 
\medskip 

The properties \eqref{eq:commCDO} and \eqref{eq:commVDO1} are in fact elementary. 
They require global conformal invariance, the tree structure of OPE diagrams
and the commutativity property \eqref{eq:commdisjoint}. To begin with, let us 
recall that we associated two disjoint sets $I_{r,1}$ and $I_{r,2}$ to every 
link. As we pointed out before, the Casimir differential operators 
\eqref{eq:Casdiffop} do not depend on whether we used the generators 
$\mathcal{T}_\alpha^{(I_{r,1})}$ or $\mathcal{T}_\alpha^{(I_{r,2})}$ to construct them. 
Let us briefly discuss 
the details of the proof. Note that we think of $\mathcal{D}$ as an operator acting 
on correlation functions $\cG$. This is signaled by the subscript $|\cG$ in the definition 
of the Casimir differential operators. In evaluating products of first order 
differential operators, one can only apply the Ward identity \eqref{eq:ward} to the 
rightmost operator, which acts directly on the correlation function, and not on some derivative thereof. But once we have converted the rightmost operators 
$\mathcal{T}^{(I_{r,1})}$ into $-\mathcal{T}^{(I_{r,2})}$, they will commute with all 
operators to their left, such that we can freely move them all the way to the left and proceed to 
apply the Ward identity to the next set of first order operators, and so on. If we finally 
take into account that the invariants $\kappa$ of the conformal Lie algebra are 
symmetric, we arrive at an expression for the Casimir differential operators in 
terms of $\mathcal{T}^{(I_{r,2})}$. 

A similar analysis can be carried out for vertex operators, see also Subsection~\ref{sec:3pt}. 
Without loss of generality we can assume that we have constructed our 
vertex operators in terms of the generators $\mathcal{T}^{(I_{\rho,1})}$ and 
$\mathcal{T}^{(I_{\rho,2})}$ and want to switch to constructing them from 
$\mathcal{T}^{(I_{\rho,1})}$ and $\mathcal{T}^{(I_{\rho,3})}$ instead. To do 
so we make use of the invariance condition  
\begin{equation} \label{eq:T213} 
\left[\mathcal{T}^{(I_{\rho,2})}_{\alpha}\right]_{|\cG} = - 
\left[\mathcal{T}^{(I_{\rho,1})}_{\alpha} - 
\mathcal{T}^{(I_{\rho,3})}_{\alpha}\right]_{|\cG}\ 
\end{equation} 
that follows from relations \eqref{eq:NIrho} and \eqref{eq:ward}. After we
apply this to the rightmost operator in the vertex differential operator, we use 
the commutativity property \eqref{eq:commdisjoint} to move the generators 
$\mathcal{T}^{(I_{\rho,1})}$ and $\mathcal{T}^{(I_{\rho,3})}$ to the left of 
$\mathcal{T}^{(I_{\rho,2})}$. We continue this replacement process until all 
the generators $\mathcal{T}^{(I_{\rho,2})}$ are removed and using symmetry 
of the tensor $\kappa$ we find 
\begin{equation} 
 \mathcal{D}^{p,\nu}_{\rho,12}
= (-1)^{p-\nu} \sum_{\mu=0}^{p-\nu} \binom{p-\nu}{\mu} 
  \mathcal{D}^{p,\nu+\mu}_{\rho,13}
\end{equation} 
along with a similar relation for the Pfaffian vertex operators for $p=d/2+1$ and $d$ even. 
Since the last term in this sum with $\mu = p-\nu$ is just a Casimir operator, we 
have managed to express all vertex differential operators that are constructed from 
the generators associated with $I_{\rho,1}$ and $I_{\rho,2}$ as a linear combination 
of the vertex operators associated with the pair $I_{\rho,1}$ and 
$I_{\rho,3}$ and a Casimir operator. This is the main input in proving the 
commutativity statements \eqref{eq:commCDO} and \eqref{eq:commVDO1}. 
\medskip  
 
Let us start with a pair of links $r,r'$. Each of these links divides the 
set of external points into the two disjoint sets $I_{r,1}, I_{r,2}$ and 
$I_{r',1}, I_{r',2}$ respectively. Since the OPE diagram is a tree, it is always possible 
to find a pair $i,j = 1,2$ such that $I_{r,i} \cap I_{r',j} = \emptyset$. 
For this choice 
\begin{equation}
[\, \mathcal{D}^p_r\,, \, \mathcal{D}^q_{r'} \, ] = 
[\, \mathcal{D}^p_{r,i}\,, \, \mathcal{D}^q_{r',j} \, ]
 = 0 
\end{equation}
because of the commutativity property \eqref{eq:commdisjoint}. This proves 
our first claim. Note that the same arguments also apply to the case in which $r=r'$ and also if one or both operators are Pfaffian, i.e. if $d$ is 
even and $p,q = d/2+1$. 

\begin{figure}[thb]
\centering
\includegraphics{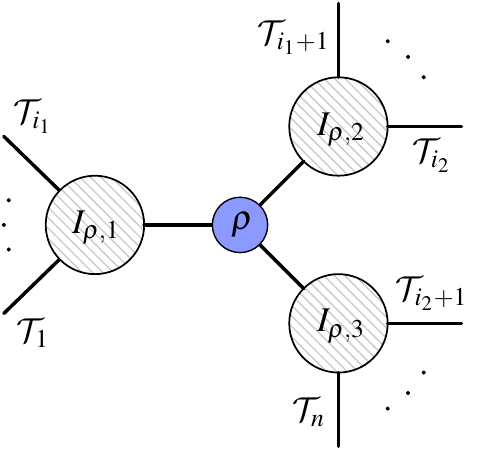}
\caption{Schematic representation of a generic OPE diagram with focus on one vertex. The choice of a 
vertex automatically divides the diagram into three branches.}
\label{fig:VertexBranches}
\end{figure}

Let us now extend this argument to include vertex differential operators. 
In order to prove that the Casimir operators associated to a link $r$ 
commute with the vertex operators associated to any vertex $\rho$ we 
recall that any choice of a vertex $\rho$ on an OPE diagram divides the 
diagram into the three distinct branches that are glued to the vertex, and we denote these by $I_{\rho,j}$ as in Figure~\ref{fig:VertexBranches}. A 
quick glance at Figure~\ref{fig:VertexBranches} suffices to conclude that given $r$ and $\rho$ 
it is possible to find a pair $i\in {1,2}$ and $j \in {1,2,3}$ such 
that $I_{r,i} \subset I_{\rho,j}$, since the link $r$ must be in one 
of the three branches. It follows that $I_{r,i} \cap \left(I_{\rho,j_1} \cup 
I_{\rho,j_2}\right) = \emptyset$ for $j_1 \neq j \neq j_2$. Commutativity of 
Casimir and vertex differential operators then follows since we 
can construct the Casimir differential operators in terms of the 
generators for $I_{r,i}$ while using the generators for $I_{\rho,j_1}$
and $I_{\rho,j_2}$ for the vertex differential operators. 

\begin{figure}[thb]
\centering
\includegraphics{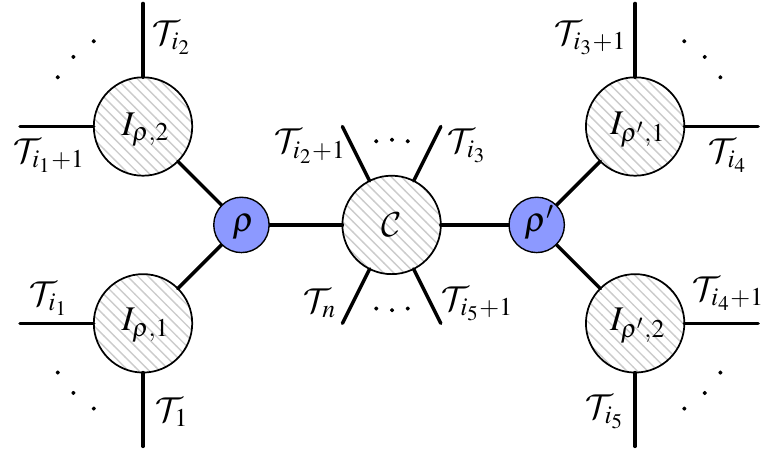}
\caption{Schematic representation of a generic OPE diagram with focus on two internal 
vertices. Operators supported around distinct vertices trivially commute, as they can be 
written in terms of generators that belong to different branches.}
\label{fig:CommutationDifferentVertices}
\end{figure}

Let us finally consider any two distinct vertices $\rho$ and $\rho^{\prime}$ on an OPE diagram. 
As we highlight in Figure~\ref{fig:CommutationDifferentVertices}, any configuration of two 
vertices divides the diagram into five parts: four external branches $I_{\rho,j}$, $I_{\rho^{\prime},j}$, 
$j=1,2$, attached to only one vertex, and the central part $\mathcal{C}$ of the diagram that is 
attached to both vertices $\rho$ and $\rho^{\prime}$. Following what we just claimed with focus on 
one vertex, we can use diagonal conformal 
symmetry to rewrite the operators around $\rho$ and $\rho^{\prime}$ to depend on disjoint 
sets of legs $I_{\rho,j}$, $I_{\rho^{\prime},j}$ with $j=1,2$. Since generators associated 
to these sets commute~\eqref{eq:commdisjoint}, it follows automatically that operators constructed 
around different vertices must commute as well.

This implies that to prove commutativity of our set of operators, we can just focus on operators 
that live around one single vertex. To prove the commutativity of these vertex operators we will now make 
use of the integrability technology that is provided by Gaudin models.

\subsection{The vertex system and Gaudin models}
\label{sec:3pt}

In this section, we will explain how the operators \eqref{eq:vdo} associated with a vertex $\rho$ 
in the OPE diagram naturally arise from a specific Gaudin model, which in particular will provide 
us with a proof of their commutativity. Let us start by reviewing briefly how Gaudin models are 
defined~\cite{Gaudin_76a,Gaudin_book83}. They are integrable systems naturally constructed from a 
choice of a simple Lie algebra $\g$. Having in mind applications of these systems to conformal field 
theories, we will choose $\g$ to be the conformal Lie algebra $\mathfrak{so}(d+1,1)$ of the Euclidean 
space $\mathbb{R}^d$, with basis $T_\alpha$ as in the previous section. The Gaudin model depends in 
general on $M$ complex numbers $w_j$, called its sites, to which are attached $M$ independent 
representations of the algebra $\g$. To obtain the vertex system we address in this section, 
we restrict our attention here to the case $M=3$ and associate with these three sites the 
representations of $\g$ corresponding to the three fields attached to the vertex $\rho$. More 
precisely, using the notation defined in Section~\ref{sec:sum} and in particular the partition 
$\underline{N} = I_{\rho,1} \cupdot I_{\rho,2} \cupdot I_{\rho,3}$ constructed from the vertex 
$\rho$, we will attach to the three sites $w_j$, $j=1,2,3$, of the Gaudin model the generators 
$\mathcal{T}_\alpha^{(I_{\rho,j})}$ which define representations of $\g$ in terms of first-order 
differential operators in the insertion points $x_i$.

A key ingredient in the construction of the Gaudin model is its so-called Lax matrix, whose 
components in the basis $T^\alpha$ are defined here as
\begin{equation}\label{eq:Lax3pt}
\Lc^\rho_\alpha(z) = \sum_{j=1}^3 \frac{\mathcal{T}_\alpha^{(I_{\rho,j})}}{z-w_j},
\end{equation}
where $z$ is an auxiliary complex variable called the spectral parameter. In the above equation, 
we have denoted the Lax matrix as $\Lc^\rho_\alpha(z)$ to emphasize that this is the matrix 
corresponding to the vertex $\rho$. For any elementary symmetric invariant tensor $\kappa_p$ 
of degree $p$ on $\g$, there is a corresponding $z$-dependent Gaudin Hamiltonian of the form
\begin{eqnarray}\label{eq:GaudinHam}
\Hc^{(p)}_\rho(z) & = & \kappa_p^{\alpha_1\cdots \alpha_p} \Lc_{\alpha_1}^\rho(z) \cdots 
\Lc_{\alpha_p}^\rho(z) + \dots,
\end{eqnarray}
where $\dots$ represent quantum corrections, involving a smaller number of components of the Lax 
matrix $\Lc_\alpha^\rho$ and their derivatives with respect to $z$. These corrections are chosen specifically to ensure that the Gaudin Hamiltonians commute for all values of the 
spectral parameter and all degrees:
\begin{equation}\label{eq:comH}
\bigl[ \Hc_\rho^{(p)}(z), \Hc_\rho^{(q)}(w) \bigr] = 0, \qquad \forall\, z,w\in\mathbb{C}, \quad  \forall\,p,q\ .
\end{equation}
The existence of such commuting Hamiltonians was first proven in~\cite{Feigin:1994in}, using some 
previously established results~\cite{Feigin:1991wy} on the so-called Feigin-Frenkel center of 
affine algebras at the critical level. The explicit expression for the quantum corrections was 
obtained in~\cite{Talalaev:2004qi,Chervov:2006xk} for Lie algebras of type A and in~\cite{Molev:2013} 
for types B, C, D, which is the case we are concerned with here (indeed, $\g=\mathfrak{so}(d+1,1)$ is 
of type B for $d$ odd and of type D for $d$ even). We refer to~\cite{Molev:2018} for a summary of
these results. The properties of the Feigin-Frenkel center were further studied in the recent 
work~\cite{Yakimova:2019}, the results of which imply that the quantum corrections in 
$\Hc^{(p)}_\rho(z)$ are sums of terms of the form
\begin{equation}\label{eq:Correc}
\tau^{\alpha_1\cdots \alpha_q} \; 
\partial_z^{r_1-1}\Lc^\rho_{\alpha_1}(z)\cdots\partial_z^{r_q-1}\Lc^\rho_{\alpha_q}(z),
\end{equation}
where $q<p$, $\tau^{\alpha_1\cdots \alpha_q}$ is a completely symmetric invariant tensor of degree 
$q$ on $\g$ and $r_1,\cdots,r_q$ are positive integers such that $r_1+\cdots+r_q=p$.  For what follows, 
it will be useful to consider the leading part of the Hamiltonian \eqref{eq:GaudinHam} alone, without 
quantum corrections, which we will denote as
\begin{equation}\label{eq:HamNoCorrec}
\widehat{\Hc}^{(p)}_\rho(z) = \kappa^{\alpha_1\cdots \alpha_p} \Lc_{\alpha_1}^\rho(z) \cdots \Lc_{\alpha_p}^\rho(z)\ .
\end{equation}
Let us finally note that the quantum corrections are absent for both the quadratic Hamiltonian
$\Hc^{(2)}_\rho(z)$ and the Pfaffian Hamiltonian $\Hc^{(d/2+1)}_\rho(z)$ that exists for $d$ even, such that these two Hamiltonians coincide with their leading parts. 

To make the link with the vertex operators defined in Section~\ref{sec:sum}, we will make a 
specific choice of the parameters $w_j$ of the Gaudin model. More precisely, we set
\begin{equation}\label{eq:pos3sites}
w_1 = 0, \qquad w_2 = 1 \qquad \text{ and } \qquad w_3 = \infty\ .
\end{equation}
In particular, the Lax matrix \eqref{eq:Lax3pt} reduces to
\begin{equation}\label{eq:LaxVertex}
\Lc^\rho_\alpha(z) = \frac{\mathcal{T}_\alpha^{(I_{\rho,1})}}{z} + \frac{\mathcal{T}_\alpha^{(I_{\rho,2})}}{z-1}\ .
\end{equation}
Let us now study the Gaudin Hamiltonians $\Hc^{(p)}_\rho(z)$ for this particular choice of 
parameters. We will first focus on their leading part $\widehat\Hc^{(p)}_\rho(z)$. Reinserting 
the above expression of the Lax matrix in eq. \eqref{eq:HamNoCorrec}, we simply find
\begin{equation}\label{eq:HamD}
\widehat{\Hc}^{(p)}_\rho(z) = \sum_{\nu=0}^p 
\binom{p}{\nu}\, \frac{\mathcal{D}_{\rho,12}^{p,\nu}}{z^\nu(z-1)^{p-\nu}},
\end{equation}
where $\mathcal{D}_{\rho,12}^{p,\nu}$ is the vertex operator defined in eq. \eqref{eq:vdo}. To obtain 
this expression, we have used the fact that $\mathcal{T}_\alpha^{(I_{\rho,1})}$ and 
$\mathcal{T}_\beta^{(I_{\rho,2})}$ commute to bring all $\mathcal{T}_\alpha^{(I_{\rho,1})}$'s to 
the left, as well as the symmetry of the tensor $\kappa_p^{\alpha_1\cdots\alpha_p}$ to relabel 
the Lie algebra indices as in eq. \eqref{eq:vdo}. Noting that the fractions $z^{-\nu}(z-1)^{\nu-p}$ 
for $\nu=0,\cdots,p$ are linearly independent functions of $z$, it is then clear that one can 
extract all the vertex operators $\mathcal{D}_{\rho,12}^{p,\nu}$ from $\widehat{\Hc}^{(p)}_\rho(z)$. 
Note that the ``extremal'' operators $\mathcal{D}_{\rho,12}^{p,0}$ and $\mathcal{D}_{\rho,12}^{p,p}$ 
coincide with the Casimir operators of the fields at the branches $I_{\rho,1}$ and $I_{\rho,2}$ of 
the vertex. The same equation also holds for the Pfaffian operators $\Hc^{(d/2+1)}_\rho(z)$.

Our goal in this section is to prove the commutativity of the vertex operators $\mathcal{D}_{\rho,12}^{p,\nu}$ 
using known results on Gaudin models. This would follow automatically from the commutativity 
\eqref{eq:comH} of the Gaudin Hamiltonians $\Hc_\rho^{(p)}(z)$ if these Hamiltonians contained the
operators $\mathcal{D}_{\rho,12}^{p,\nu}$. But we have already proven above that the latter are naturally 
extracted from the leading parts $\widehat{\Hc}^{(p)}_\rho(z)$ of the Gaudin Hamiltonians, without the 
quantum corrections. These quantum corrections are in general crucial for the commutativity of the Hamiltonians. However, we shall prove below that the quantum corrections of the specific Gaudin model considered here can always be expressed in terms of lower-degree Hamiltonians, and can thus be discarded without breaking the commutativity property. In this case, the non-corrected Hamiltonians $\widehat{\Hc}^{(p)}_\rho(z)$ pairwise commute for all values of the spectral 
parameter and all degrees, thus demonstrating the desired commutativity of the vertex operators 
$\mathcal{D}_{\rho,12}^{p,\nu}$.

Let us then analyze these quantum corrections. Recall that they are composed of terms of the form 
\eqref{eq:Correc}. Reinserting the expression \eqref{eq:LaxVertex} of the Lax matrix for the present 
choice of parameters $w_j$ in this equation, we find that the quantum corrections contain only terms 
of the form
\begin{equation}
\tau^{\alpha_1\cdots\alpha_q} \; \mathcal{T}_{\alpha_1}^{(I_{\rho,1})} \cdots 
\mathcal{T}_{\alpha_\nu}^{(I_{\rho,1})} 
\mathcal{T}_{\alpha_{\nu+1}}^{(I_{\rho,2})} \cdots \mathcal{T}_{\alpha_q}^{(I_{\rho,2})},
\end{equation}
with prefactors composed of powers of $z$ and $z-1$, and where $\tau^{\alpha_1\cdots\alpha_q}$ is a completely 
symmetric invariant tensor on $\g$ of degree $q<p$, as in eq. \eqref{eq:Correc}. In particular, 
$\tau^{\alpha_1\cdots\alpha_q}$ decomposes as a product of elementary symmetric invariant tensors $\kappa_k$,
symmetrized over the indices $\alpha_i$, with $k \leq q < p$. The correction in the above equation can thus be 
re-expressed as an algebraic combination of lower-degree vertex operators $\mathcal{D}_{\rho,12}^{k,\nu}$. Since 
these are the coefficients of the non-corrected Hamiltonian $\widehat{\Hc}^{(k)}_\rho(z)$,
recursion on the degrees shows that the quantum corrections can indeed be expressed in terms of lower-degree 
Hamiltonians, as anticipated. 
\smallskip 

Let us end this subsection with a brief discussion on the role played by the choice of labeling of the branches 
$I_{\rho,1}$, $I_{\rho,2}$ and $I_{\rho,3}$ attached to the vertex $\rho$. As mentioned in Section~\ref{sec:sum}, this choice is arbitrary but enters the definition \eqref{eq:vdo} of the vertex 
operators $\mathcal{D}_{\rho,12}^{p,\nu}$, 
which in this case contain only the generators $\mathcal{T}_{\alpha}^{(I_{\rho,1})}$ and 
$\mathcal{T}_{\alpha}^{(I_{\rho,2})}$. In the context of the 3-sites Gaudin model considered in this subsection, 
this is related to the choice of positions $w_j$ of the sites made in eq. \eqref{eq:pos3sites}. In particular, 
the absence of generators $\mathcal{T}_{\alpha}^{(I_{\rho,3})}$ in the Gaudin Hamiltonians $\Hc^{(p)}(z)$ is due 
to the fact that we sent the site $w_3$ to infinity. One could have made another choice of labeling and constructed vertex operators $\mathcal{D}_{\rho,23}^{p,\nu}$ from the generators
$\mathcal{T}_{\alpha}^{(I_{\rho,3})}$ and $\mathcal{T}_{\alpha}^{(I_{\rho,2})}$, for instance. The corresponding 
choice of positions of the sites would then be related to the initial one by the M\"obius transformation that 
exchanges $0$ and $\infty$ and fixes $1$, i.e. the inversion $z\mapsto \frac{1}{z}$. More precisely, 
under such a transformation of the spectral parameter, the Lax matrix of the Gaudin model behaves as a 1-form 
on the Riemann sphere and satisfies
\begin{equation}
-\frac{1}{z^2} \Lc_\alpha^\rho\left( \frac{1}{z} \right) = 
-\frac{\mathcal{T}_{\alpha}^{(I_{\rho,1})}+\mathcal{T}_{\alpha}^{(I_{\rho,2})}}{z} + 
\frac{\mathcal{T}_{\alpha}^{(I_{\rho,2})}}{z-1}\ .
\end{equation}
Acting on the correlation function $\cG_N$, which satisfies the Ward identities \eqref{eq:ward}, this Lax 
matrix then becomes
\begin{equation}
-\frac{1}{z^2} \Lc_\alpha^\rho \left(\frac{1}{z} \right)_{|\cG} = 
\left[\frac{\mathcal{T}_{\alpha}^{(I_{\rho,3})}}{z} + 
\frac{\mathcal{T}_{\alpha}^{(I_{\rho,2})}}{z-1}\right]_{|\cG},
\end{equation}
and thus coincides with the Lax matrix from which one would build the vertex operators 
$\mathcal{D}_{\rho,23}^{p,\nu}$ with generators $\mathcal{T}_{\alpha}^{(I_{\rho,3})}$ and 
$\mathcal{T}_{\alpha}^{(I_{\rho,2})}$. This proves that the operators $\mathcal{D}_{\rho,23}^{p,\nu}$ 
can naturally be extracted from the generating functions
\begin{equation}\label{eq:Ham1overz}
\frac{(-1)^p}{z^{2p}} \widehat\Hc_\rho^{(p)}\left( \frac{1}{z} \right)_{|\cG}\ . 
\end{equation}
Using the expression \eqref{eq:HamD} of $\widehat\Hc_\rho^{(p)}\left( z \right)$ in terms of the initial 
vertex operators $\mathcal{D}_{\rho,12}^{p,\nu}$, we thus get that the $\mathcal{D}_{\rho,23}^{p,\nu}$'s 
are linear combinations of the $\mathcal{D}_{\rho,12}^{p,\nu}$'s, as was demonstrated by direct computation 
in the previous subsection. Let us note that the use of the Ward identities was a crucial step in the above 
reasoning, as highlighted for instance by the subscript $\cG$ in eq. \eqref{eq:Ham1overz}. This step should 
be performed with care, in particular when using the Ward identities 
to replace $-\mathcal{T}_{\alpha}^{(I_{\rho,1})} 
-\mathcal{T}_{\alpha}^{(I_{\rho,2})}$ by $\mathcal{T}_{\alpha}^{(I_{\rho,3})}$ in the 
Hamiltonians  \eqref{eq:Ham1overz}. Indeed, one can use the Ward identities only for generators on the 
right. In order 
to do so, one thus has to commute generators to bring them to the right, replace them through 
the Ward identities and commute them back to their original place. Although this procedure can in general create 
non-trivial corrections, it can in fact be done freely in the case at hand: indeed, commuting operators 
$\mathcal{T}_{\alpha_k}^{(I_{\rho,j})}$ and $\mathcal{T}_{\alpha_l}^{(I_{\rho,j})}$ within the Hamiltonian 
$\widehat \Hc_\rho^{(p)}$ creates a term proportional to the structure constant $f_{\alpha_k\alpha_l}^
{\hspace{17pt}\beta}$, which vanishes when contracted with the symmetric tensor $\kappa_p^{\alpha_1\cdots
\alpha_p}$. This ensures that the Hamiltonian \eqref{eq:Ham1overz} indeed serves as a generating function 
of the operators $\mathcal{D}_{\rho,23}^{p,\nu}$ built from $\mathcal{T}_{\alpha}^{(I_{\rho,3})}$ 
and $\mathcal{T}_{\alpha}^{(I_{\rho,2})}$. A similar reasoning applies for the other choices of labeling, 
by considering the appropriate M\"obius transformations that permute the sites $0$, 
$1$ and $\infty$ of our 3-site Gaudin model. 

\subsection{Restricted vertices and relations between vertex operators} 
\label{sec:dependences}

In the previous subsection we have shown that all of the operators listed in 
eq.\ \eqref{eq:vdo} commute with each other. As we have pointed out before, we 
did not include operators with $\nu =0$ and $\nu =p$ in the list since these 
coincide with Casimir differential operators,  
\begin{equation}
\mathcal{D}^{p,0}_{\rho,12} = \mathcal{D}^p_{r_1} \quad , \quad 
\mathcal{D}^{p,p}_{\rho,12} = \mathcal{D}^p_{r_2}\ . 
\end{equation} 
Here $r_i$ denotes the link that is attached to the $i^{th}$ leg of the 
vertex $\rho$, i.e. for which $I_{r_i,j} = I_{\rho,i}$ with either $j=1$ 
or $j=2$. The remaining $p-1$ operators satisfy one more linear relation 
since 
\begin{equation} 
\sum_{\nu= 0}^p \binom{p}{\nu} \, \mathcal{D}^{p,\nu}_\rho = \mathcal{D}^p_{r_3} \ . 
\end{equation} 
Let us note that this relation also applies to the Pfaffian vertex operators 
that exist for $p=d/2+1$ when $d$ is even. We have used this relation to drop 
one of the vertex differential operators. Once these obvious relations are 
taken care of, the total number of commuting vertex differential operators 
is given by eq.\ \eqref{eq:novertfull} and matches precisely the maximal 
number of cross ratios that can be associated to a single (generic) vertex, 
see upper bound of eq.\ \eqref{eq:novertrest} in the introduction. But restricted vertices carry fewer variables, so their corresponding differential operators (constructed in the previous section) must obey further relations. It is the main goal of this subsection to discuss these 
relations. We will also check that, once these are taken into account, the 
number of remaining vertex differential operators matches the number 
\eqref{eq:novertrest} of cross ratios at restricted vertices. 
\medskip 

Our arguments are based on an important auxiliary result concerning the 
differential operators $\mathcal{T}^{(I)}_\alpha$ that are associated to 
some subset $I \subset \underline N$ of order $|I| \leq N/2$. To present 
this requires a bit of preparation. Up to this point there was no need to 
spell out the precise form of the symmetric invariant tensors $\kappa_p$
that we used to construct our differential operators. Now we need to be a 
bit more specific. As is well known, such tensors can be realized as 
symmetrized traces,  
\begin{equation} 
\kappa^{\alpha_1 \cdots \alpha_p}_p = \textit{tr} 
\left( T^{(\, \alpha_1} \cdots T^{\alpha_p \,)}\right) 
= \textit{str} 
\left( T^{\alpha_1} \cdots T^{\alpha_p}\right)\ . 
\end{equation} 
Here $T^\alpha$ denote the generators of the conformal Lie algebra and 
$( \dots )$ signal symmetrization with respect to the indices. In the 
following we shall use the symbol \textit{str} to denote this symmetrized 
trace. The trace can be taken in any faithful representation. The simplest 
of such choices is to use the fundamental representation. In order to 
construct the associated symmetric invariants more explicitly, we shall 
replace the index $\alpha$ that enumerates the basis of the conformal algebra 
by a pair $\alpha = [AB]$ where $A,B$ run through $A,B = 0,1, \dots, 
d+1$ with $T^{[AB]} = - T^{[BA]}$. In the fundamental representation, 
the matrix elements of these generators take the form 
$$ \bigl( T^{[AB]}_{f} \bigr)^{C}_{\;\;D} = \eta^{AC}\,  \delta^B_{\ D} - \eta^{BC}\,  \delta^A_{\ D} \ , $$
where $\eta_{AB}$ is the Minkowski metric with signature $(d+1,1)$ and $\eta^{AB}$ is its inverse. This makes it now easy to compute $\kappa_p$ explicitly. The only issue 
arises in even $d$. In this case the symmetrized traces in the fundamental 
representation do not generate all the invariants. In order to obtain the 
missing invariant, one has to include the trace in a chiral representation. 
The standard construction employs the spinor representation in which 
generators $T^{[AB]}$ are represented as 
$$ \bigl( T^{[AB]}_{s} \bigr)^{\sigma}_{\;\;\tau} =  
\frac14\,  [\, \gamma^A\, ,\, \gamma^B\, ]^{\sigma}_{\;\;\tau}  \ ,$$ 
where $\gamma^A$ are the $d+2$-dimensional $\gamma$ matrices and the 
matrix indices are $\sigma,\tau = 1, \dots, 2^{d/2+1}$. One can then project to 
a chiral spinor representation with the help of $\gamma_c \sim \gamma^0 
\cdots \gamma^{d+1}$.  
\smallskip 

Let us now introduce the symbol $\mathcal{T}^{(I)}$ to denote the 
following Lie-algebra valued differential operators 
$$ \mathcal{T}^{(I)} = T^\alpha \, \cdot \,  \mathcal{T}^{(I)}_\alpha = 
\frac 12 T^{[AB]} \, \cdot \, \mathcal{T}^{(I)}_{[AB]} \ . $$  
Upon evaluation in some finite-dimensional representation, such as the 
fundamental or the spinor representation, these become matrix valued 
differential operators. With this notation we write our set 
\eqref{eq:vdo} as 
\begin{equation}
    \mathcal{D}^{p,\nu}_{\rho,12}  =   
    \textit{str}_f\left(\underbrace{\mathcal{T}^{(I_{\rho,1})}...\mathcal{T}^{(I_{\rho,1})}}_{\nu}
    \underbrace{\mathcal{T}^{(I_{\rho,2})}...\mathcal{T}^{(I_{\rho,2})}}_{p-\nu}\right)_{|\cG} 
    \label{eq:vdoalt} 
\end{equation}
when $d$ is odd and the parameters $p$ and $\nu$ assume the values 
$p=2,4,\dots, d+1$ and $\nu = 1, \dots, p-1$, as usual. For even dimension 
$d$, on the other hand, we use the symmetrized trace in the fundamental 
representation for $p=2,4, \dots, d$ and construct the missing Pfaffian 
vertex differential operators as 
\begin{equation}
    \mathcal{D}^{d/2+1,\nu}_{\rho,12}  =   
    \textit{str}_s\left(\underbrace{\mathcal{T}^{(I_{\rho,1})}...\mathcal{T}^{(I_{\rho,1})}}_{\nu}\underbrace{
    \mathcal{T}^{(I_{\rho,2})}...\mathcal{T}^{(I_{\rho,2})}}_{d/2+1-\nu}\gamma_c\right)_{|\cG} 
    \label{eq:pdoalt} 
\end{equation}
where we take the trace in the spinor representation and include the factor 
$\gamma_c$ in the argument. 
\medskip  
 
In finding relations between the vertex differential operators for restricted 
vertices we actually work with the total symbols of the differential operators 
rather than the operators themselves. This means that we replace the partial 
derivatives $\partial^{(i)}_\mu$ with commuting coordinates $p^i_\mu$. The 
associated matrices of functions of $x_i^\mu$ and $p^i_\mu$ will be denoted by $\bar{\mathcal{T}}^{(I)}$. As before we shall add a subscript $f,s$ to denote 
the matrices in the fundamental and the spinor representation. After passing 
to the total symbol the entries of the matrices commute and we can drop the 
symmetrization prescription when taking traces. As a result, the total symbols of the 
vertex differential operators are simply traces of powers of the matrices 
$\bar{\mathcal{T}}^{(I)}$. 

We are now ready to state the main result needed to elucidate the relations between vertex differential operators. It concerns the matrix elements of the $n^\textit{th}$ 
power of the matrices $\bar{\mathcal{T}}^{(I)}_{f,s}$ 
for the fundamental and the spinor representations. In both cases, these 
matrix elements are functions of $x_i^\mu$ and $p_\mu^i$ with $i \in I$. 
Our main claim is that these matrix elements can be expressed in terms of 
lower order ones of the same form whenever $n > 2\dep_I$, where $\dep_I = \dep(I,d)$ 
is the integer defined in eq.\ \eqref{eq:depth}. More precisely, for each matrix 
element $AB$ there exist coefficients $\varrho^{(n,m)}_{AB}$ such that 
\begin{equation}
\label{eq:lemmaf}
    \bigl(\bar{\mathcal{T}}^{(I)\, n}_f\bigr)\null^{A}_{\;\;B} = \sum_{m=0}^{2\dep_I} 
  \varrho^{(n,m)}_{f;AB}\, \bigl(\bar{\mathcal{T}}^{(I)\, m}_f\bigr)\null^{A}_{\;\;B}\ .  
\end{equation} 
Note that there is no summation over $A,B$ on the right hand side. The coefficients
$\varrho_{AB}$ depend only on the external conformal weights $\Delta_i$ and the 
total symbols $\bar{\mathcal{D}}^p_{(I)}$ of the Casimir operators associated 
with the index set $I$. If the index set $I$ has depth $\dep_I=1$, for example, i.e.
\ if the $\bar{\mathcal{T}}^{(I)}$ describe the action of the conformal algebra on 
a single scalar primary, then starting from $n=3$ all matrix elements can be expressed 
in terms of lower order ones. In the case of the spinor representation one has a 
very similar relation
\begin{equation}
\label{eq:lemmas}
  \bigl(\bar{\mathcal{T}}^{(I)\, n}_s\bigr)\null^{\sigma}_{\;\;\tau} = \sum_{m=0}^{\dep_I} 
  \varrho^{(n, m)}_{s;\sigma,\tau}\, \bigl(\bar{\mathcal{T}}^{(I)\, m}_s\bigr)\null^{\sigma}_{\;\;\tau}\ . 
\end{equation} 
which now applies for $n > \dep_I$ and involves a summation over $m$ that ends 
at $\dep_I = \dep(I,d)$, see definition \eqref{eq:depth}. So if $\dep_I=1$, for 
example, the matrix elements of the square are expressible in terms of the matrix 
elements of $\bar{\mathcal{T}}^{(I)}_s$. We verify both statements
\eqref{eq:lemmaf} and \eqref{eq:lemmas} in Appendix~\ref{appendix:prooflemma} using embedding 
space formalism, see Appendix~\ref{appendix:Embeddingspace}. 
\medskip  
 
We are now prepared to discuss relations between vertex differential operators. 
Let us consider a vertex $\rho$ inside our OPE diagram. As we have explained before, 
$\rho$ splits the set $\underline{N}$ into three subsets $I_{\rho,i}$ with $i=1,2,3$. 
Each of these sets determines an integer $\dep_i = \dep(I_{\rho,i},d)$. Let us 
suppose that we construct the vertex differential operators using $\mathcal{T}
^{(I_{\rho,i})}$ for $i=1,2$ as in eq.\ \eqref{eq:vdoalt}. 
If one of the integers $\dep_1$ or $\dep_2$ is smaller than $r_d$ we immediately obtain 
relations among the vertex differential operators. In fact, when applied to the matrices $\bar{\mathcal{T}}^{(I_\rho,1)}$, our claim 
\eqref{eq:lemmaf} implies that all operators we obtain when $\nu > 2 \dep_1$ can be expressed in terms of Casimir and 
vertex differential operators of lower order. The same is true when $p-\nu > 2\dep_2$, 
as follows again from eq.\ \eqref{eq:lemmaf}, but this time applied to 
$\bar {\mathcal{T}}^{(I_\rho,2)}$. Consequently, for any $p \geq 2\dep_1,2\dep_2$, we 
can restrict the range of the index $\nu$ to be $p-2\dep_2 \leq \nu \leq 2\dep_1$, 
with $\nu = 0,p$ excluded as before.  
 
\begin{figure}[thb]
\centering
\includegraphics{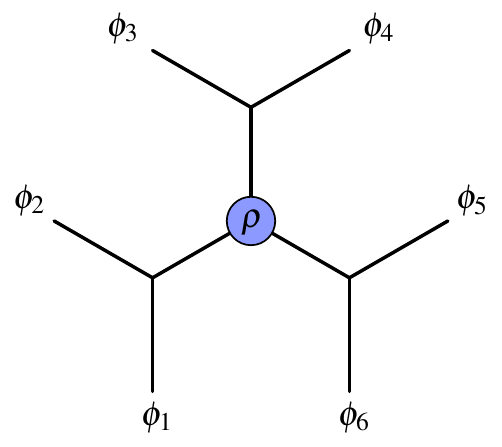}
\caption{Snowflake OPE diagram. Here the tensor product of any two branches around 
the vertex $\rho$ would allow a mixed symmetry tensor of depth $\dep = 4$ to appear in $d
\ge7$, but diagonal conformal symmetry constrains this to match with the symmetric 
traceless tensor produced on the third branch.}
\label{fig:Snowflake}
\end{figure}

But this does not yet include the full set of relations that appears whenever $\dep_3$ 
is smaller than $\textit{min}(\dep_1+\dep_2,r_d)$. One of the simplest examples of this 
occurs in the 6-point snowflake channel in $d>3$, see Figure~\ref{fig:Snowflake}, 
where two symmetric traceless tensor operators on two branches are combined at the central 
vertex $\rho$ to form another symmetric traceless tensor on the third branch.

To take restrictions from the third branch $I_3$ into account, it is sufficient to 
use the total symbol of eq.\ \eqref{eq:T213} and impose once more our dependence statement \eqref{eq:lemmaf} 
for product matrices, this time for $\mathcal{T}^{(I_{\rho,3})}$. This tells us that the 
matrix elements of powers 
\begin{equation}
    \left(\bar{\mathcal{T}}^{(I_{\rho,1})}+\bar{\mathcal{T}}^{(I_{\rho,2})}
    \right)^{n\,\,A}_{\hspace{14pt} B}
    \label{binomialpowertensorp}
\end{equation}
with $n > 2\dep_3$ can be written in terms of lower order terms. In order to convert this 
observation into relations among the vertex differential operators of order $p$, we can 
multiply the expression \eqref{binomialpowertensorp} with some appropriate powers of 
$\bar{\mathcal{T}}^{(I_{\rho,i})}$, $i=1,2$, and consider the matrix elements of the 
products
\begin{equation}
    \left(\bar{\mathcal{T}}^{(I_{\rho,1})}+\bar{\mathcal{T}}^{(I_{\rho,2})}\right)^{n} 
    \left(\bar{\mathcal{T}}^{(I_{\rho,2})}\right)^{\nu} 
    \left(\bar{\mathcal{T}}^{(I_{\rho,1})}\right)^{p-n-\nu} 
    \label{binomialpowertensorpfill}
\end{equation}
for $n > 2 \dep_3$, any allowed value of $p > n$ and $\nu = 0, \dots, p-n$. By binomial 
expansion, we can write the expression \eqref{binomialpowertensorpfill} as a linear
combination of our basic vertex differential operators. After taking relations on the
branches $I_{\rho,1}$ and $I_{\rho,2}$ into account, we obtain an additional nontrivial 
relation from the third branch $I_{\rho,3}$.  For example, in the cases where $n$ is odd, 
contracting~\eqref{binomialpowertensorp} with $\bigl(\mathcal{T}^{(I_{\rho,2})}\bigr)^B_{\;\;A}$ 
leads to a relation between vertex and Casimir operators of order $n+1$ and further lower order 
operators. This effectively reduces the amount of vertex operators at order $p$ by up 
to $p-2\dep_3-1$, though the actual number can be lower in case there are less than $p-
2\dep_3-1$ vertex operators at order $p$ left after imposing the constraints from the 
first two branches. If we are interested in counting the number of vertex operators at a given even order $p$, the procedure we just outlined is summarized in the following counting formula
\begin{equation}
    n_{vdo,\rho}^{(p)}=\textit{max}\left[\left((p-2)-\sum_{i=1}^3\Theta_0(p-2\dep_i)(p-2\dep_i-1)\right),0\right]\,,
    \label{eq:countingnormal}
\end{equation}
where $\Theta_0$ is the Heaviside step function with $\Theta_0(0)=0$, the factor $(p-2)$ gives the maximal amount of vertex operators at order $p$, the factor $(p-2\dep_i-1)$ corresponds to the number of relations introduced for every $\dep_i< p/2$, and the maximum enforces the number to be $0$ if there are more than $(p-2)$ relations in total.

Our description of relations between vertex differential operators exploited the 
auxiliary statement \eqref{eq:lemmaf} and did not include the Pfaffian vertex 
differential operators. It is clear, however, that precisely the same reasoning 
also applies to the latter using eq. \eqref{eq:lemmas} instead of eq. \eqref{eq:lemmaf}. The counting formula \eqref{eq:countingnormal} gets also slightly modified for this Pfaffian case
\begin{equation}
    n_{vdo,\rho}^{(p=d/2+1)}=\textit{max}\left[\left((p-2)-\sum_{i=1}^3\Theta_0(p-\dep_i)(p-\dep_i-1)\right),0\right]\,.
    \label{eq:countingPfaffians}
\end{equation}
Note that the arguments we have outlined here exhibit relations between vertex 
operators, but we have not shown that these relations are complete, i.e. that the 
remaining vertex differential operators are in fact independent. A priori, it could 
in fact happen that the exceeding relations we get from here, or additional relations 
obtained from a different reasoning, could provide additional dependencies. We checked however that summing the counting formulas \eqref{eq:countingnormal} and \eqref{eq:countingPfaffians} for all allowed orders $p$ in a given dimension $d$, gives rise to a number of vertex differential operators equal to the number of cross ratios~\eqref{eq:novertrest} associated with every allowed vertex.
This provides strong evidence in favor of the independence of our vertex differential operators. In some particularly relevant cases in lower dimensions we can also prove independence. 
\medskip 
 
\textit{Example: The $N=6$ snowflake channel for $d=7$.} Let us see how all of this works 
in the example of a snowflake channel in $d=7$, presented in Figure~\ref{fig:Snowflake}.
We enumerate the internal links by $r=1,2,3$. The associated index sets $I_{r,i}$ are  
$I_{1,1} = \{1,2\}, I_{2,1} = \{3,4\}, \dots$. Here we have two symmetric traceless 
tensors associated with $r=1,2$. In a more general OPE diagram these could produce a mixed 
symmetry tensor with maximal depth $\dep = r_d = 4$, but in the snowflake diagram the field 
on the third link must also be a symmetric traceless tensor of depth $\dep(I_3,d=7) = 2$. 
Our prescription tells us to consider operators~\eqref{eq:vdoalt} up to $p=8$. Eliminating 
powers of $\mathcal{T}_1 = \mathcal{T}^{(12)}$ and $\mathcal{T}_2 = \mathcal{T}^{(34)}$ 
higher than 4, it immediately follows that there are no vertex operators of order 8
\begin{equation}
    \cancel{\mathcal{T}_{1}^7\mathcal{T}_{2}}\,, \quad 
    \cancel{\mathcal{T}_{1}^6\mathcal{T}_{2}^2}\,, \quad 
    \cancel{\mathcal{T}_{1}^5\mathcal{T}_{2}^3}\,, \quad 
    \cancel{\mathcal{T}_{1}^3\mathcal{T}_{2}^5}\,, \quad 
    \cancel{\mathcal{T}_{1}^2\mathcal{T}_{2}^6}\,, \quad 
    \cancel{\mathcal{T}_{1}\mathcal{T}_{2}^7}\,,
\end{equation}
while there could be up to two operators of order 6
\begin{equation}
    \cancel{\mathcal{T}_{1}^5\mathcal{T}_{2}}\,, \quad 
    \mathcal{T}_{1}^4\mathcal{T}_{2}^2\,, \quad 
    \mathcal{T}_{1}^2\mathcal{T}_{2}^4\,, \quad \
    \cancel{\mathcal{T}_{1}\mathcal{T}_{2}^5}\,,
    \label{sixthordermonomialssnowflake}
\end{equation}
and two operators of order 4
\begin{equation}
    \mathcal{T}_{1}^3\mathcal{T}_{2}\,, \quad 
    \mathcal{T}_{1}\mathcal{T}_{2}^3\,.
    \label{fourthordermonomialssnowflake}
\end{equation}
Here and in the following steps we are using notation for which stroked terms are 
dependent on lower order operators. Let us also recall that the operators with $\nu = p/2$ 
have been omitted to account for the relation between the vertex and Casimir differential 
operators for the third leg. The reduction of $\mathcal{T}_3 = \mathcal{T}^{(56)}$ to a
symmetric traceless tensor implies the existence of $p-2\dep_{3}-1$ relations between 
$p$ order monomials. The only useful relation in this case is the one produced for 
$p=6$, coming from the expansion
\begin{equation}
    \cancel{\left(\mathcal{T}_{1}+\mathcal{T}_{2} \right)^5}=
    \mathcal{T}_{1}^5+5\mathcal{T}_{1}^4\mathcal{T}_{2}+
    10\mathcal{T}_{1}^3\mathcal{T}_{2}^2+10\mathcal{T}_{1}^2\mathcal{T}_{2}^3+
    5\mathcal{T}_{1}\mathcal{T}_{2}^4+\mathcal{T}_{2}^5\,,
\end{equation}
which can be contracted with either $\mathcal{T}_{1}$ or $\mathcal{T}_{2}$ and 
traced over to get a relation between the sixth order monomials (including 
the one associated to the Casimir of the third leg):
\begin{equation}
    \cancel{\left(\mathcal{T}_{1}+\mathcal{T}_{2} \right)^{5}\mathcal{T}_{2}}=
    \cancel{\mathcal{T}_{1}^5\mathcal{T}_{2}}+
    5\mathcal{T}_{1}^4\mathcal{T}_{2}^2+10\mathcal{T}_{1}^3\mathcal{T}_{2}^3+
    10\mathcal{T}_{1}^2\mathcal{T}_{2}^4+5\cancel{\mathcal{T}_{1}\mathcal{T}_{2}^5}+
    \cancel{\mathcal{T}_{2}^6}\,.
\end{equation}
This reduces the amount of independent vertex operators by one, bringing us to a total 
of three independent operators, which matches with the number of associated cross 
ratios.

\section{OPE channels and limits of Gaudin models}
\label{sec:GaudinOPE}
 
At this point we have defined a set of differential operators associated with the intermediate 
fields and the individual vertices of a given OPE diagram with $N$ external fields. The new vertex operators were constructed in Subsection~\ref{sec:3pt} from a Gaudin model with three sites, which was crucial in proving their commutativity. Our construction of the vertex 
operators has been \textit{local} in its focus on a particular building block, 
namely a single vertex that is associated to a local element of the OPE diagram. The purpose 
of this section is to adopt a more \textit{global} perspective by showing that the whole set of Casimir and vertex differential operators for any $N$-point OPE channel can be obtained by taking an appropriate limit of an $N$-site Gaudin model. The
$N$-site Gaudin model itself makes no reference to the choice of OPE channel. The latter enters only 
through the choice of limit. We will therefore refer to these limits as OPE limits of 
the $N$-site Gaudin model. These OPE limits are generalizations of the so-called bending flow 
and caterpillar limits that have been considered in the mathematical literature, see e.g. 
\cite{Chervov:2007dn,Chervov:2009,rybnikov2016cactus}. The same limit of a $4$-site Gaudin 
model - which may be identified with the elliptic Inozemtsev model \cite{Argyres:2021iws} -  
has also appeared in the physics literature recently \cite{Eberhardt:2020ewh}. 
Eberhardt, Komatsu and Mizera have shown that the limit theory coincides with the hyperbolic Calogero-Sutherland model, which is known to describe the Casimir equations of 
$4$-point conformal blocks. 

\subsection{\texorpdfstring{$N$}{N} sites Gaudin model and OPE limits}
\label{sec:GaudinLim}

Let us first define the Gaudin model that we will use in this section. Since this construction 
is similar to the one of the 3-sites Gaudin model considered in Section~\ref{sec:3pt}, we 
refer to that section for details and references. As before, we consider a Gaudin model based 
on the conformal Lie algebra $\g=\mathfrak{so}(d+1,1)$ but now with $N$ sites, whose positions 
$w_1,\cdots,w_N\in\mathbb{C}$ are for the moment arbitrary. We naturally associate these sites 
with the $N$ external fields in the correlation function under consideration and more precisely 
attach to each site $i\in\lbrace 1,\cdots,N\rbrace$ the representation of $\g$ defined by 
the generators $\mathcal{T}_\alpha^{(i)}$, which describe the action of the conformal 
transformations on the scalar field $\phi_i(x_i)$ in terms of first-order differential 
operators. Then we define the (components of the) Lax matrix of the model as
\begin{equation} \label{eq:LaxN} 
\Lc_\alpha(z,w_i) = \sum_{i=1}^N \frac{\mathcal{T}_\alpha^{(i)}}{z-w_i}\ .
\end{equation}
The associated Gaudin Hamiltonians $\mathcal{H}^{(p)}(z,w_i)$ of degree $p$ are given by the 
same equation \eqref{eq:GaudinHam} that we used for the 3-site case. Recall that $\kappa_p$ denotes the conformally invariant symmetric tensor of degree $p$ and $\dots$ represent quantum corrections. 
The latter have the same form as in the 3-site case, see eq.\ \eqref{eq:Correc}. It is well 
known that these $N$-site Gaudin Hamiltonians commute, just as their three site analogues, i.e.\ 
they satisfy eq.\ \eqref{eq:comH}. At the same time, it is easy to verify that they are invariant 
under diagonal conformal transformations, 
\begin{equation}\label{eq:comDiag}
\bigl[ \Hc^{(p)}(z,w_i), \mathcal{T}^{(\underline{N})}_\alpha \bigr] = 0, \qquad \forall\, z\in\mathbb{C}, \quad \forall\,p,
\end{equation}
where $\mathcal{T}^{(\underline{N})}_\alpha = \sum_{i=1}^N \mathcal{T}^{(i)}_\alpha$ are the 
diagonal conformal generators that also appear in the Ward identities \eqref{eq:ward}. This 
means that Gaudin Hamiltonians descend to correlation functions $\cG$.  

The Hamiltonian $\mathcal{H}^{(p)}(z,w_i)$ of the $N$-site Gaudin model depends on the $N$ 
complex parameters $w_i$ that specify the poles of the Lax matrix. These parameters 
have no a priori interpretation in the context of correlation functions. Note that we can always apply M\"obius transformations on the $z$-variable to fix three of the $w_i$. This is what allowed us in the previous section to set the parameters of the 3-site Gaudin model to the specific values \eqref{eq:pos3sites}, in which case the Lax matrix \eqref{eq:LaxVertex} and its corresponding differential operators contain no extra parameters. Our main claim is that we can 
reconstruct the entire set of 3-site Lax matrices $\mathcal{L}^\rho_\alpha(z)$, as well as the associated vertex Hamiltonians $\mathcal{H}^{(p)}_\rho(z)$, one set 
for each vertex $\rho$, from the $N$-site Lax matrix and the associated Gaudin 
Hamiltonians $\mathcal{H}^{(p)}(z) = \mathcal{H}^{(p)}(z,w_i)$ by taking appropriate 
scaled limits of the complex parameters $w_i$.% 
\medskip

In order to make a precise statement, we need a bit of preparation. The limits we are 
about to discuss must depend on the choice of the OPE channel. So let us assume we are 
given such a channel $\mathcal{C}$. In order to define the limits we pick an (arbitrary) 
external edge in the diagram, which will serve as a reference point and which, up to 
reordering, we can suppose to have label $N$. As this edge is external, it is attached 
to a unique vertex, which we will denote by $\rho_\ast$. Such a choice of reference 
vertex defines a so-called rooted tree representation of the diagram. We then draw the 
OPE diagram on a plane, with the vertex $\rho_\ast$ situated at the top and with each vertex 
having two downward edges attached. Such a representation on a plane forces us to make 
a choice of which edges are pointing towards the left and which edges are pointing 
towards the right: this choice is arbitrary, and gives rise to what is called a plane 
(or ordered) representation of the underlying rooted tree. We give an example of 
such a plane rooted tree representation for an 8-point OPE diagram in Figure 
\ref{fig:diag} below.
\begin{figure}[H]
\begin{center}
\includegraphics[scale=1]{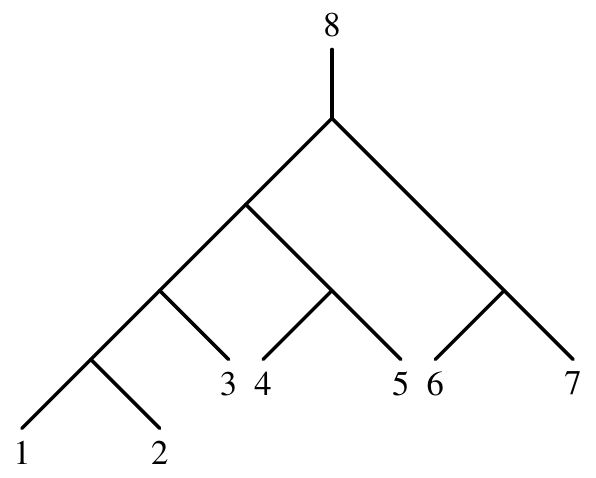}
\caption{Plane rooted tree representation of an OPE diagram with 8 external fields.}\label{fig:diag}
\end{center}
\end{figure}
Recall from Section~\ref{sec:sum} that each vertex $\rho$ of the OPE diagram defines 
a partition $\underline{N} = I_{\rho,1} \cupdot I_{\rho,2} \cupdot I_{\rho,3}$, with 
the sets $I_{\rho,j}$ formed by the labels of the external fields attached to the three 
branches of the vertex. Although the choice of labeling of these branches was arbitrary 
in Section~\ref{sec:sum}, we will now fix it using the plane rooted tree representation 
of the diagram picked above: choose the branch $I_{\rho,1}$ to be the 
one pointing to the bottom left and the branch $I_{\rho,2}$ to be the one pointing to the 
bottom right. By construction, the last branch $I_{\rho,3}$ then always points to the top 
and contains the reference point $N$. 
\medskip 

Each vertex $\rho$ in the diagram is thereby associated with a sequence $s^\rho = 
(s^\rho_1,s^\rho_2, \dots, s^\rho_{n_\rho})$ of elements $s^\rho_a \in \{1,2\}$. This sequence 
$s^\rho$ encodes the path from $\rho_\ast$ to $\rho$. It tells us whether we have to 
move to the left (for $s_a^\rho=1$) or right (for $s_a^\rho=2$) every time we reach a new vertex until we arrive at $\rho$ after 
$n_\rho$ steps. We shall also refer to the length $n_\rho$ of the sequence as the depth of the vertex and to $s^\rho$ as the \textit{binary sequence} of $\rho$. Note 
that the top vertex $\rho_\ast$ has depth $n_{\rho_\ast}=0$. Let us point out that 
the notion of depth used in this section refers to the distance from the root $\rho_\ast$ and is very different from the depth $\dep$ introduced 
in eq.\ \eqref{eq:depth} of the  introduction. 

In order to construct the limit of the Gaudin model that we are interested in, we will need to assign a polynomial $g_\rho(\varpi)$
to each vertex $\rho$. If $s^\rho$ is the binary sequence associated with the vertex $\rho$, 
the polynomial $g_\rho$ is defined as 
\begin{equation} \label{eq:grho}
g_\rho(\varpi) = \sum_{a=1}^{n_\rho} \varpi^{a-1} \delta_{s_a^\rho,2} \ . 
\end{equation} 
Obviously the top vertex $\rho_\ast$ is assigned to $g_{\rho_\ast}(\varpi) = 0$. 
The vertices of depth $n_\rho = 1$ are associated with $g_{\rho_1}(\varpi)= 0$ or 
$g_{\rho_2}(\varpi) = 1$, depending on whether they are reached from $\rho_\ast$
by going down to the left ($\rho_1$) or to the right ($\rho_2$). 

Similarly, we can assign polynomials $f_i(\varpi)$ to each external edge $1 \leq i<N$ at 
the bottom of the plane rooted tree. Once again, we can encode the path from $\rho_\ast$ 
down to the edge $i$ by a binary sequence $s^i = (s^i_1, s^i_2,  \dots, s_{n_i}^i)$. 
The length $n_i$ of the sequence $s^i$ is also referred to as the depth of the edge $i$. 
Now we introduce 
\begin{equation} \label{eq:fi} 
f_i(\varpi) = \sum_{a=1}^{n_i} \varpi^{a-1} \delta_{s^i_a,2} 
+ \varpi^{n_i}\delta_{s^i_{n_i},1}\ .  
\end{equation} 
and set 
\begin{equation}  \label{eq:wHighest} 
f_N(\varpi) = \varpi^{-1}
\end{equation} 
for the external edge of the reference field at the top of the plane 
rooted tree. Thereby we have now set up all the necessary notation that is needed to 
construct the relevant scaling limits of the $N$-site Gaudin model. 
\medskip 

We can now move on to the main result of this section, namely how to reconstruct
the vertex Hamiltonians $\mathcal{H}^{(p)}_\rho(z)$ of the 3-site Gaudin model of the previous section from the $N$-site Hamiltonians $\mathcal{H}^{(p)}(z) =
\mathcal{H}^{(p)}(z,w_i)$. To this end, we will first construct the vertex Lax matrices
\eqref{eq:LaxVertex} from the Lax matrix \eqref{eq:LaxN} before studying the
associated Hamiltonians \eqref{eq:GaudinHam} in the limit. As it turns out, we 
can recover the parameter free Lax matrix $\Lc^\rho$ that is associated with  
the vertex $\rho$ as 
\begin{equation}\label{eq:LimitL}
\Lc_\alpha^\rho(z) = \frac{\mathcal{T}_\alpha^{(I_{\rho,1})}}{z} + 
\frac{\mathcal{T}_\alpha^{(I_{\rho,2})}}{z-1} = \lim_{\varpi \to 0}\  
\varpi^{n_\rho} \Lc_\alpha \bigl( \varpi^{n_\rho}z + g_\rho(\varpi), 
w_i = f_i(\varpi)\bigr)\ .
\end{equation}
Let us note that in the limit, the site $w_N= \varpi^{-1}$ associated with the reference 
field $N$ goes to infinity, while the sites of the other external fields approach $z=0$
or $z=1$ depending on whether they are located at the right or left branch of the the 
plane rooted tree, i.e.\ whether their binary sequence $s^\rho$ starts with $s^\rho_1 = 
1$ or $s^\rho_1 =2$. We shall prove eq.\ \eqref{eq:LimitL} in the third Subsection~\ref{sec:recursion} through 
a recursive procedure that will also offer insight into the construction of the polynomials
$g_\rho$ and $f_i$. 
\medskip 

Let us now turn to the limit construction for the Gaudin Hamiltonians. We claim that the Hamiltonians $\mathcal{H}^{(p)}(z,w_i)$ of the $N$-sites Gaudin model give rise to the Hamiltonians $\mathcal{H}^{(p)}_\rho(z)$ 
of the different 3-sites vertex Gaudin models defined in Section~\ref{sec:3pt} as 
\begin{equation}\label{eq:LimitH}
\Hc_\rho^{(p)}(z) = \lim_{\varpi \to 0} \varpi^{pn_\rho} 
\Hc^{(p)} \bigl( \varpi^{n_\rho}z + g_\rho(\varpi),w_i = f_i(\varpi) \bigr)\ .
\end{equation}
The fact that this statement holds for the leading part of the Hamiltonians, without quantum corrections, 
follows directly from the corresponding limit \eqref{eq:LimitL} of the Lax matrix
\begin{equation*}
\varpi^{pn_\rho}\kappa_p^{\alpha_1\cdots \alpha_p} \Lc_{\alpha_1}\bigl( \varpi^{n_\rho}z + g_\rho(\varpi) \bigr) 
\cdots \Lc_{\alpha_p}\bigl( \varpi^{n_\rho}z + g_\rho(\varpi) \bigr) \xrightarrow{\varpi \to 0} 
\kappa^{\alpha_1\cdots \alpha_p} \Lc_{\alpha_1}^\rho(z) \cdots \Lc_{\alpha_p}^\rho(z)\ .
\end{equation*}
But it requires a bit of work to argue that the quantum corrections also have the required behaviour 
under the limit. Consider a term of the form \eqref{eq:Correc} in the correction: by appropriately distributing the powers $\varpi^{pn_\rho}$, using the fact that $r_1 + \cdots + r_q = p$, and 
performing the change of spectral parameter $z\mapsto z'_\rho(\varpi) =  \varpi^{n_\rho}z + 
g_\rho(\varpi)$ in the derivatives, we find that
\begin{align*}
&\varpi^{pn_\rho}\tau^{\alpha_1\cdots \alpha_q} \; \partial_{z'_\rho(\varpi)}^{r_1-1}\Lc_{\alpha_1}
\bigl(z'_\rho(\varpi) \bigr)\cdots\partial_{z'_\rho(\varpi)}^{r_q-1}\Lc_{\alpha_q}\bigl(z'_\rho(\varpi) \bigr)\\
& \hspace{15pt} = \hspace{15pt}\tau^{\alpha_1\cdots\alpha_q} \partial^{r_1-1}_z \Bigl( \omega^{n_\rho} 
\Lc_{\alpha_1} \bigl(\varpi^{n_\rho}z + g_\rho(\varpi) \bigr) \Bigr) \cdots \partial^{r_q-1}_z 
\Bigl( \omega^{n_\rho} \Lc_{\alpha_q} \bigl(\varpi^{n_\rho}z + g_\rho(\varpi) \bigr) \Bigr) \\
&\hspace{7pt} \xrightarrow{\varpi \to 0}\hspace{7pt}  \tau^{\alpha_1\cdots \alpha_q} \;
\partial_z^{r_1-1}\Lc^\rho_{\alpha_1}(z)\cdots\partial_z^{r_q-1}\Lc^\rho_{\alpha_q}(z),
\end{align*}
such that this correction term reduces in the OPE limit to the corresponding correction in $\Hc_\rho^{(p)}(z)$.

As the vertex operators $\mathcal{D}_\rho^{p,\nu}$ and the Casimir operators $\mathcal{D}_r^p$ of the 
intermediate fields attached to the vertex $\rho$ are naturally extracted from the Hamiltonian 
$\Hc^{(p)}_\rho(z)$, the property \eqref{eq:LimitH} shows that the full set of operators defined in 
Section~\ref{sec:sum} can be obtained from the limit of the $N$-sites Gaudin model considered here. Before the limit $\varpi\to 0$, The 
commutativity property \eqref{eq:comH} of the $N$-site Gaudin Hamiltonians can be written as
\begin{equation}
\Bigl[ \Hc^{(p)}\bigl(\varpi^{n_\rho}z + g_\rho(\varpi) \bigr), \Hc^{(q)}\bigl(\varpi^{n_\rho}w + 
g_\rho(\varpi) \bigr) \Bigr] = 0, \qquad \forall\, z,w\in\mathbb{C}, \quad \forall\,p,q, \quad \forall\,\rho,\rho'.
\end{equation}
for arbitrary $\varpi$, and is therefore preserved in the limit $\varpi\to 0$,
\begin{equation}
\bigl[ \Hc^{(p)}_\rho(z), \Hc^{(q)}_{\rho'}(w) \bigr] = 0, \qquad 
\forall\, z,w\in\mathbb{C}, \quad \forall\,p,q, \quad \forall\,\rho,\rho'.
\end{equation}
This provides an alternative proof of the commutativity of all Casimir operators $\mathcal{D}_r^p$ 
and vertex operators $\mathcal{D}_\rho^{p,\nu}$. Moreover, this statement now holds without needing to use conformal Ward identities. The proof relies on a specific choice of labeling of the edges at vertices, given by a plane rooted tree representation of the OPE 
diagram. Different such representations of the diagram correspond to different limits of the same 
underlying $N$-sites Gaudin model and give rise to different sets of commuting operators, which 
however generate the same algebra when acting on solutions of the conformal Ward identities. Finally, let
us note that the above construction automatically ensures the compatibility of these operators with the conformal Ward identities, since taking appropriate limits of eq.
\eqref{eq:comDiag} demonstrates that they commute with the diagonal conformal generators 
$\mathcal{T}_\alpha^{(\underline{N})}$.

\subsection{Examples}

Before we prove our main result, let us illustrate the construction of the operators from limits of Gaudin models with 
two examples. The first one addresses the so-called comb channel OPE diagrams for which we 
have already outlined the limit in \cite{Buric:2020dyz}. The second example deals with
the snowflake OPE channel of the $N=6$-point function.  

\paragraph{Comb channel.} Let us consider the comb channel OPE diagram with $N$ external fields. 
To apply the construction of the present section, we first need to pick a plane rooted tree 
representation of this diagram. We will choose to represent it with all internal edges 
pointing towards the bottom left. We then label the external edges of the tree as follows: we 
let $N$ be the top edge of the tree, $1$ be edge furthest to the left and label by $2,
\cdots,N-1$ the external edges pointing to the bottom right at each vertex, from the bottom 
to the top. Moreover, following our conventions in \cite{Buric:2020dyz}, we enumerate the vertices $\rho = [r]$ by an integer $r=1, \dots, N-2$, from bottom to top. We represent this plane rooted tree in Figure \ref{fig:comb} below, with external edges indicated 
in black and vertices in blue.
\begin{figure}[H]
\begin{center}
\includegraphics[scale=1]{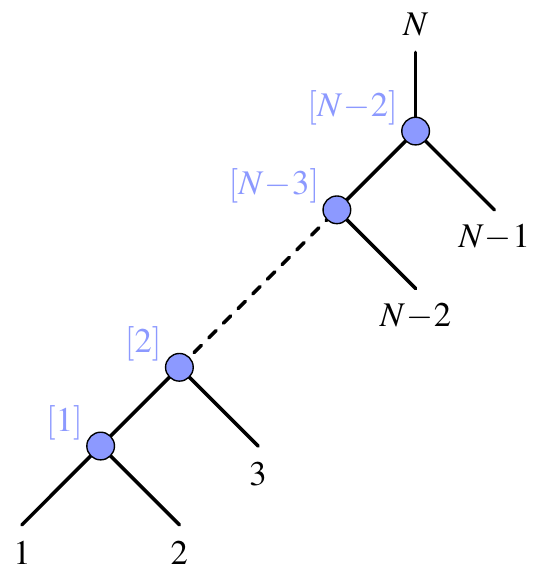}
\caption{Choice of plane rooted tree representation of the comb channel OPE diagram 
with $N$ points.}\label{fig:comb}
\end{center}
\end{figure}
One can compute the limit of the Gaudin model associated with this tree using the construction outlined in the previous subsection. For the polynomials $f_i$ that determine the parameters 
$w_i$ of the Gaudin model, one finds from eq. \eqref{eq:fi} that 
\begin{equation}
w_i = f_i(\varpi) = \varpi^{N-1-i}, \qquad \forall\,i\in\lbrace 1,\cdots,N \rbrace\ .
\end{equation}
Let us now consider the vertices $\rho = [r]$, with $r=1, \dots, N-2$. From the general construction, and formula
\eqref{eq:grho} in particular, we find
\begin{equation}
n_{[r]} = N-2-r \qquad \text{ and } \qquad g_{[r]}(\varpi) = 0, \qquad 
\forall\, r\in\lbrace 1,\cdots,N-2 \rbrace\ .
\end{equation}
Note in particular that for this choice of plane rooted tree, the polynomial functions 
$g_{\rho}$ are all zero, since all vertices $\rho = [r]$ sit on the most left branch 
of the tree. The limit of the Gaudin Lax matrix
\begin{equation}
\Lc_\alpha(z,w_i=f_i(\varpi)) = \sum_{i=1}^N \frac{\mathcal{T}^{(i)}_\alpha}{z-\varpi^{N-1-i}}
\end{equation}
associated with the vertex $[r]$ then reads
\begin{equation}
\varpi^{N-2-r} \Lc_\alpha \bigl( \varpi^{N-2-r} z \bigr) \xrightarrow{\varpi\to 0} \Lc_\alpha^{[r]}(z) = \frac{\mathcal{T}_\alpha^{(1)}+\cdots+\mathcal{T}_\alpha^{(r)}}{z} + \frac{\mathcal{T}^{(r+1)}}{z-1}.
\end{equation}
In sum, the vertex Gaudin Hamiltonians of the comb channel OPE limit are
\begin{equation}
\varpi^{p(N-2-r)} \Hc^{(p)} \bigl( \varpi^{N-2-r} z \bigr) \xrightarrow{\varpi\to 0} \Hc_{[r]}^{(p)}(z).
\end{equation}
The above limit coincides exactly with the one introduced in~\cite{Buric:2020dyz} to describe the comb 
channel, thus showing that the results of~\cite{Buric:2020dyz} are contained in the more general 
construction discussed here.
\medskip 

\begin{figure}[H]
\begin{center}
\includegraphics[scale=1]{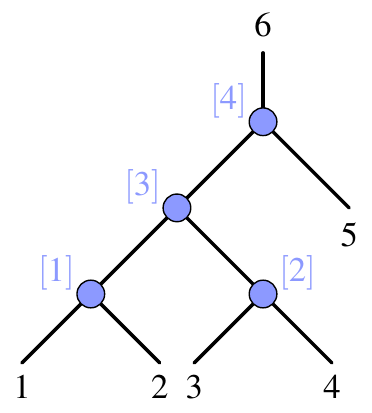}
\caption{Choice of plane rooted tree representation of the snowflake OPE diagram.}\label{fig:snowflake}
\end{center}
\end{figure}

\paragraph{Snowflake channel.} The results of the present article allow us to discuss more general 
topologies of OPE diagrams than the comb channel. The first example of such a topology is the snowflake 
channel of $6$-point functions. We represent this OPE diagram as a plane rooted tree following the 
conventions of Figure \ref{fig:snowflake}, where the external edges are labeled in black from $1$ 
to $6$ and the vertices are labeled in blue from $[1]$ to $[4]$. Note in particular that the 
internal vertex of the diagram corresponds here to the label $[3]$.
We can immediately read off the depth of the four vertices as
\begin{equation}
n_{[1]}=n_{[2]}=2, \qquad n_{[3]}=1, \qquad n_{[4]}=0.
\end{equation}
We can now encode the positions of all four vertices in a binary sequence and apply the formulas 
\eqref{eq:grho} to construct the polynomials $g_\rho$, yielding  
\begin{equation}
g_{[1]}(\varpi) = g_{[3]}(\varpi) = g_{[4]}(\varpi) = 0 \qquad \text{ and } \qquad g_{[2]}(\varpi) = \varpi.
\end{equation}
Similarly, one can encode the five external edges at the bottom of the diagram in binary sequences and apply eq. \eqref{eq:fi} to determine the positions of the $N=6$ sites, 
\begin{equation}\label{Eq:zSnow}
w_1 = \varpi^3, \qquad w_2 = \varpi^2, \qquad w_3 = \varpi+\varpi^3, \qquad w_4 = \varpi+\varpi^2, 
\qquad w_5 = 1, \qquad w_6 = \frac{1}{\varpi}.
\end{equation}
Inserting this parameterization of the complex parameters $w_i$ in terms of $\varpi$ back into the Lax 
matrix of the 6-sites Gaudin model, we obtain 
\begin{equation}
\Lc_\alpha(z) = \frac{\mathcal{T}_\alpha^{(1)}}{z-\varpi^3} + 
\frac{\mathcal{T}_\alpha^{(2)}}{z-\varpi^2} + 
\frac{\mathcal{T}_\alpha^{(3)}}{z-\varpi-\varpi^3} + 
\frac{\mathcal{T}_\alpha^{(4)}}{z-\varpi-\varpi^2} + 
\frac{\mathcal{T}_\alpha^{(5)}}{z-1} + \frac{\varpi\,\mathcal{T}_\alpha^{(6)}}{\varpi\,z-1}.
\end{equation}
Given this expression and our formulas for $n_\rho$ and $g_\rho$, it is now straightforward to check
the limits \eqref{eq:LimitL} for all four vertices, 
\begin{equation}
\varpi^2 \Lc_\alpha(\varpi^2 z) \xrightarrow{\varpi\to 0} \frac{\mathcal{T}_\alpha^{(1)}}{z} +
\frac{\mathcal{T}_\alpha^{(2)}}{z-1}, \qquad \varpi^2 \Lc_\alpha(\varpi^2 z+\varpi) 
\xrightarrow{\varpi\to 0} \frac{\mathcal{T}_\alpha^{(3)}}{z} + \frac{\mathcal{T}_\alpha^{(4)}}{z-1},
\end{equation}
\begin{equation*}
\varpi \Lc_\alpha(\varpi z) \xrightarrow{\varpi\to 0} 
\frac{\mathcal{T}_\alpha^{(1)}+\mathcal{T}_\alpha^{(2)}}{z} + 
\frac{\mathcal{T}_\alpha^{(3)}+\mathcal{T}_\alpha^{(4)}}{z-1}, \qquad \Lc_\alpha(z) 
\xrightarrow{\varpi\to 0} \frac{\mathcal{T}_\alpha^{(1)}+\mathcal{T}_\alpha^{(2)}+
\mathcal{T}_\alpha^{(3)}+\mathcal{T}_\alpha^{(4)}}{z} + \frac{\mathcal{T}_\alpha^{(5)}}{z-1}.
\end{equation*}
These indeed give the expected vertex Lax matrices $\Lc_\alpha^\rho = \Lc_\alpha^{[r]}$ 
for the vertices labeled by $r=1,2,3,4$. These two examples suffice to gain 
a first intuition into how we take limits of Gaudin models and thereby manage to 
embed the vertex Lax matrices into the full $N$-sites model. We will now explain the derivation of our results for general OPE diagrams.

\subsection{Recursive proof of the limits}
\label{sec:recursion}

\paragraph{Subtrees.} Our goal in this subsection is to prove that our limit construction is 
indeed able to recover all vertex Lax matrices, as we have claimed. Let us consider 
some OPE channel $\mathcal{C}$ represented by a plane rooted tree $T$. The approach that we follow is recursive. Let us consider the top vertex 
$\rho_\ast$ of the tree, which is by construction attached to the external edge $N$. We 
denote by $e'$ and $e''$ the left and right downward edges attached to $\rho_\ast$ (which can 
correspond to either external or intermediate fields depending on the topology of the 
diagram). We can then see the tree $T$ as being composed of the vertex $\rho_\ast$ and of two 
(plane rooted) subtrees with reference edges $e'$ and $e''$, which we will call $T'$ 
and $T''$ respectively. In Figure \ref{fig:diag2} below, we illustrate the two subtrees 
obtained in the example of Figure \ref{fig:diag}.
\begin{figure}[H]
\begin{center}
\includegraphics[scale=1]{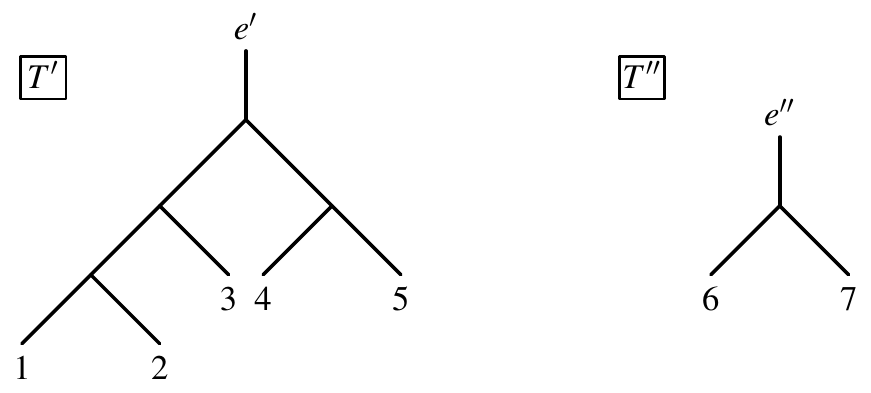}
\caption{Subtrees of the tree \ref{fig:diag}.}\label{fig:diag2}
\end{center}
\end{figure}
We will now prove that if the limit construction of Subsection~\ref{sec:GaudinLim} holds for the subtrees $T'$ and $T''$, then it also holds for the initial tree $T$, thus proving that it holds for any tree by induction. Let us first introduce 
some useful notation. We will denote by $E'$ and $E''$ the external edges of $T$ that belong to 
the subtrees $T'$ and $T''$ respectively. Note that if $T'$ is not trivial, i.e. if 
$e'$ is not an external edge of the initial tree $T$, then the full set of external edges of $T'$ 
is $E' \cupdot \lbrace e'\rbrace$ (since $e'$ is external in $T'$ but not in $T$). On the 
other hand, if $e'$ is external in $T$, then we simply have $E'=\lbrace e' \rbrace$ and the 
subtree $T'$ is trivial. Let us also denote by $V$, $V'$ and $V''$ the set of vertices of 
$T$, $T'$ and $T''$, such that $V = V' \cupdot V'' \cupdot \lbrace \rho_\ast \rbrace$.

\paragraph{Recursion relations.} Recall the polynomial $g_\rho(\varpi)$ defined in eq. \eqref{eq:grho} for any vertex $\rho\in V$ of $T$ in terms of the binary sequence $s^\rho=(s^\rho_1,\cdots,s^\rho_{n_\rho})$. Let us suppose that this vertex is contained in the subtree $T'$ and thus belongs to $V'$: it is then associated with a binary sequence $s'^{\,\rho}=(s'^{\,\rho}_1,\cdots,s'^{\,\rho}_{n'_\rho})$ in $T'$. By construction, the depth $n'_\rho$ of $\rho$ in $T'$ is given by $n_\rho-1$. Moreover, it is clear that the binary sequence of $\rho$ in $T$ is related to that in $T'$ by $s^\rho = (1, s'^{\,\rho}_1,\cdots,s'^{\,\rho}_{n_\rho-1})$. 
Indeed, since $\rho$ belongs to $T'$, the path from $\rho_\ast$ to $\rho$ starts by going to the bottom left ($s^\rho_1=1$), and is then given by the path from $e'$ to $\rho$, encoded by $s'^{\,\rho}$. It is then clear that the polynomial $g_\rho(\varpi)$ defined by eq. \eqref{eq:grho} is related to the corresponding polynomial $g'_\rho(\varpi)$ defined for $T'$ by $g_\rho(\varpi)=\varpi\,g'_\rho(\varpi)$. Similarly, if $\rho$ belongs to $V''$, we have $s^\rho = (2, s''^{\,\rho}_1,\cdots,s''^{\,\rho}_{n_\rho-1})$ and thus $g_\rho(\varpi)=1+\varpi\,g''_\rho(\varpi)$. In conclusion, the polynomials $g_\rho(\varpi)$ satisfy the recursion relation
\begin{equation}\label{eq:gRec}
g_\rho(\varpi) = \left\lbrace \begin{array}{ll}
\varpi g'_\rho(\varpi) & \text{ if } \rho\in V', \\
1+\varpi g''_\rho(\varpi) \:\,& \text{ if } \rho\in V'', \\
0 & \text{ if } \rho=\rho_{\ast}.
\end{array} \right.
\end{equation}

A similar analysis can be performed for the sites $w_i=f_i(\varpi)$ associated with the external edges $i\in\underline{N}$ through eq. \eqref{eq:fi}, distinguishing three cases. If $i=N$ is the top reference edge, then we recall that $w_N=f_N(\varpi)=\varpi^{-1}$. If $i\in E'$ is an edge belonging to the subtree $T'$, then one can relate the binary sequences $s^{i}$ and $s'^{\,i}$ describing $i$ in $T$ and $T'$ in a similar way as for the vertices in the paragraph above. We then find that the polynomial $f_{i}(\varpi)$ satisfy the recursion relation
\begin{equation}\label{eq:w'}
f_{i}(\varpi) = \left\lbrace \begin{array}{ll}
\varpi & \text{ if } e' \text{ is external in } T, \\
\varpi f'_{i}(\varpi) \hspace{7pt} & \text{ else.} 
\end{array} \right. 
\end{equation}
Note that in the first case, the subtree $T'$ is trivial and the index $i$ is then necessarily equal to
$e'$, while in the second case $i$ is different from $e'$ and the subtree $T'$ is therefore non-trivial. Finally, if $i\in E''$ belongs to the subtree $T''$, then we similarly find
\begin{equation}\label{eq:w''}
f_{i}(\varpi) = \left\lbrace \begin{array}{ll}
1 & \text{ if } e'' \text{ is external in } T, \\
1 + \varpi f''_{i}(\varpi) \hspace{7pt} & \text{ else.} 
\end{array} \right. 
\end{equation}
Here, $i=e''$ in the first case and $i\neq e''$ in the second one.

\paragraph{Induction hypotheses.} We will now suppose that the limit procedure defined in Subsection~\ref{sec:GaudinLim} holds for the subtrees $T'$ and $T''$. To phrase these induction hypotheses more precisely, let us focus first on the subtree $T'$. If it is non-trivial, i.e. if $e'$ is not external in $T$, the external edges of $T'$ are $E' \cupdot \lbrace e' \rbrace$. We then introduce the Gaudin Lax matrix associated with $T'$ as
\begin{equation}\label{eq:L'}
\Lc'_\alpha(z) = \sum_{i\in E' \cupdot \lbrace e' \rbrace}  \frac{\mathcal{T}_\alpha^{(i)}}{z-f'_{i}(\varpi)},
\end{equation}
where the sites associated with the external edges $E'\cupdot \lbrace e' \rbrace$ are set to the positions $f'_{i}(\varpi)$ prescribed by the limit procedure of Subsection~\ref{sec:GaudinLim}. Here, the generators $\mathcal{T}_\alpha^{(i)}$ associated with external fields $i \in E'$ are defined by their expression in the initial tree $T$, while the generators associated with $e'$ are defined by $\mathcal{T}_\alpha^{(e')}=\mathcal{T}_\alpha^{(\underline{N}\setminus E')}$. By construction, the latter satisfy the commutation relations of $\g$ and commute with the other generators $\mathcal{T}_\alpha^{(i)}$, $i\in E'$, as required. Moreover, this definition ensures that the diagonal conformal generators $\sum_{i\in\underline{N}} \mathcal{T}_\alpha^{(i)}$ of the tree $T$ coincides with the ones $\sum_{i\in E' \cupdot \lbrace e' \rbrace} \mathcal{T}_\alpha^{(i)}$ of $T'$ (however, as we will see, the definition of $\mathcal{T}_\alpha^{(e')}$ is in fact irrelevant for the recursive proof). 

As $T'$ is assumed to be non-trivial here, its vertex set $V'$ is non-empty. The induction hypothesis that we make in this subsection is then that eq. \eqref{eq:LimitL} holds for the subtree $T'$, that is to say
\begin{equation}\label{eq:LimitL'}
\Lc_\alpha^\rho(z) = \frac{\mathcal{T}_\alpha^{(I_{\rho,1})}}{z} + 
\frac{\mathcal{T}_\alpha^{(I_{\rho,2})}}{z-1} = \lim_{\varpi \to 0}\  
\varpi^{n'_\rho} \Lc'_\alpha \bigl( \varpi^{n'_\rho}z + g'_\rho(\varpi) \bigr), \qquad \forall \rho \in V'.
\end{equation}
In this equation, we used the Lax matrix $\Lc_\alpha^\rho(z)$ associated with the vertex $\rho$ as defined in the initial tree $T$: indeed, it is clear that this vertex Lax matrix coincides with the one associated with $\rho$ in the subtree $T'$ (in particular, the subsets of external edges $I_{\rho,1}$ and $I_{\rho,2}$ associated with the left and right branches of $\rho$ are the same when defined for $T$ as when defined for $T'$).

Similarly, if $T''$ is non-trivial, we consider the associated Lax matrix
\begin{equation}\label{eq:L''}
\Lc''_\alpha(z) = \sum_{i\in E'' \cupdot \lbrace e'' \rbrace}  \frac{\mathcal{T}_\alpha^{(i)}}{z-f''_{i}(\varpi)},
\end{equation}
and suppose that it satisfies the induction hypothesis
\begin{equation}\label{eq:LimitL''}
\Lc_\alpha^\rho(z) = \frac{\mathcal{T}_\alpha^{(I_{\rho,1})}}{z} + 
\frac{\mathcal{T}_\alpha^{(I_{\rho,2})}}{z-1} = \lim_{\varpi \to 0}\  
\varpi^{n''_\rho} \Lc''_\alpha \bigl( \varpi^{n''_\rho}z + g''_\rho(\varpi) \bigr), \qquad \forall \rho \in V''.
\end{equation}

\paragraph{Proof of the induction.} We are now in a position to prove that the induction carries from the subtrees $T'$ and $T''$ to $T$. For that, we will show that the limit \eqref{eq:LimitL} holds for every vertex $\rho\in V$, with three cases to distinguish. If $\rho\in V'$ belongs to the subtree $T'$, we will use the induction hypothesis \eqref{eq:LimitL'} and the recursion relations \eqref{eq:gRec}, first case, and \eqref{eq:w'}. Similarly, if $\rho\in V''$ belongs to $T''$, we will use the induction hypothesis \eqref{eq:LimitL''} and the recursion relations \eqref{eq:gRec}, second case, and \eqref{eq:w''}. Finally, if $\rho$ is the reference vertex $\rho_\ast$, then the limit will follow without having to use any induction hypothesis. As these proofs are rather technical, we gather them in Appendix \ref{app:ProofRec}.

\section{Example: 5-point conformal blocks}
\label{section:fivepoints}

As an example of our construction of commuting differential operators, let us consider a correlator of five scalar fields
\begin{equation}
    \expval{\phi_1\phi_2\phi_3\phi_4\phi_5}
\end{equation}
and fix the OPE decomposition as in Figure~\ref{fig:FivePointsOPE}.

	This correlator can be be written schematically as
	\begin{equation}
		\expval{\phi_1\phi_2\phi_3\phi_4\phi_5}=\Omega^{(\Delta_i)}_5(x_i)\psi^{(\Delta_i)}(u_1,\dots,u_5)
		\label{formfivecorrelator}
	\end{equation}
	where $\Omega^{(\Delta_i)}_5(x_i)$ is a prefactor that takes into account the covariance of the correlator with respect to conformal transformations, while $\psi^{(\Delta_i)}(u_1,\dots,u_5)$ is a conformally invariant function which depends on five cross ratios and admits a conformal block decomposition. In order to obtain differential equations for 5-point conformal blocks, one first needs to determine which Casimir and vertex operators characterize these blocks, and then compute their action on the space of cross ratios $u_i$.
	
	For the OPE decomposition of Figure~\ref{fig:FivePointsOPE}, the recipe of Section~\ref{sect:vertexsystem} instructs us to construct four Casimir operators, two for each internal leg
	\begin{align}
		\mathcal{D}^{2}_{(12)}=&\left(\mathcal{T}_1+\mathcal{T}_2\right)_{[AB]}\left(\mathcal{T}_1+\mathcal{T}_2\right)^{[BA]}\,,\label{fivepointsQuadCasimir12}\\
		\mathcal{D}^{2}_{(45)}=&\left(\mathcal{T}_4+\mathcal{T}_5\right)_{[AB]}\left(\mathcal{T}_4+\mathcal{T}_5\right)^{[BA]}\,,\label{fivepointsQuadCasimir45}\\
		\mathcal{D}^{4}_{(12)}=&\left(\mathcal{T}_1+\mathcal{T}_2\right)_{[AB]}\left(\mathcal{T}_1+\mathcal{T}_2\right)^{[BC]}\left(\mathcal{T}_1+\mathcal{T}_2\right)_{[CD]}\left(\mathcal{T}_1+\mathcal{T}_2\right)^{[DA]}\,,\\
		\mathcal{D}^{4}_{(45)}=&\left(\mathcal{T}_4+\mathcal{T}_5\right)_{[AB]}\left(\mathcal{T}_4+\mathcal{T}_5\right)^{[BC]}\left(\mathcal{T}_4+\mathcal{T}_5\right)_{[CD]}\left(\mathcal{T}_4+\mathcal{T}_5\right)^{[DA]}\,,
	\end{align}
	and one vertex operator
	\begin{equation}
	    \mathcal{D}_{\rho,(12)3}^{4,3}=\left(\mathcal{T}_1+\mathcal{T}_2\right)_{[AB]}\left(\mathcal{T}_1+\mathcal{T}_2\right)^{[BC]}\left(\mathcal{T}_1+\mathcal{T}_2\right)_{[CD]}\left(\mathcal{T}_3\right)^{[DA]}\,.
	    \label{fivepointsvertexop}
	\end{equation}
	Note that -- in agreement with the general recipe -- the vertex operator is not uniquely defined  and~\eqref{fivepointsvertexop}  can be shifted by terms proportional to $(\mathcal{T}_1+\mathcal{T}_2)^2(\mathcal{T}_3)^2$ or to the Casimir operators.\footnote{In~\cite{Buric:2020dyz}, we had picked the more symmetric expression $(\mathcal{T}_1+\mathcal{T}_2-\mathcal{T}_3)^4$ for the vertex operator, which leads to an equivalent set of operators.}

	To make explicit computations, we will use the embedding space formalism of~\cite{Costa:2011mg}, with which one can efficiently compute the action of the differential operators in cross ratio space; we briefly review this formalism in Appendix~\ref{appendix:Embeddingspace}. Note here that the dimension of spacetime only appears in our computations as a parameter when contracting Kronecker deltas $\delta^A_A=d+2$, and can be therefore kept generic.
	
	The first step is the choice of prefactor $\Omega^{(\Delta_i)}_5$ and cross ratios $u_i$. We chose to use the same conventions as~\cite{Rosenhaus:2018zqn}, where the author computed 5-point blocks in the case of scalar exchange; the expression for $\Omega^{(\Delta_i)}_5$ in physical space coordinates can be easily translated into one in embedding space through the simple relation
	\begin{equation}
		-2 X_{i}\cdot X_{j}=(x_{i}-x_j)^2\,,
		\label{relationconversion}
	\end{equation}
	and one obtains (up to an overall normalization) the prefactor
	\begin{equation}
		\Omega^{(\Delta_i)}_5=\frac{\left(\frac{X_2\cdot X_3}{X_1\cdot X_3}\right)^{\frac{\Delta_1-\Delta_2}{2}} \left(\frac{X_2\cdot X_4}{X_2\cdot X_3}\right)^{\frac{\Delta_3}{2}} 		\left(\frac{X_3\cdot X_5}{X_3\cdot X_4}\right)^{\frac{\Delta_4-\Delta_5}{2}}}{\left(X_1\cdot X_2\right)^{\frac{\Delta_1+\Delta_2}{2}}\left( X_3\cdot X_4\right)^{\frac{\Delta_3}{2}}\left(X_4\cdot X_5\right)^{\frac{\Delta_4+\Delta_5}{2}}} \,.
		\label{Omegafivepts}
	\end{equation}
	Regarding the cross ratios, it is natural to build four of these using the same construction of the standard 4-point cross ratios $(u,v)$ introduced in~\cite{Dolan:2000ut}, but supported on two different sets of points
	\begin{equation}
		\begin{gathered}
		u_1=\frac{\left(X_1\cdot X_2\right) \left(X_3\cdot X_4\right)}{\left(X_1\cdot X_3\right) \left(X_2 \cdot X_4\right)}\,, \qquad u_2=\frac{\left(X_1\cdot X_4\right) \left(X_2\cdot X_3\right)}{\left(X_1\cdot X_3\right) \left(X_2 \cdot X_4\right)}\,,\\
		u_3=\frac{\left(X_2\cdot X_3\right) \left(X_4\cdot X_5\right)}{\left(X_2\cdot X_4\right) \left(X_3 \cdot X_5\right)}\,, \qquad u_4=\frac{\left(X_2\cdot X_5\right) \left(X_3\cdot X_4\right)}{\left(X_2\cdot X_4\right) \left(X_3 \cdot X_5\right)}\,,
		\end{gathered}
	\label{firstfourucrossratios}
	\end{equation}
	while an interesting choice for the fifth cross ratio is
	\begin{equation}
	    u_5=\frac{\left(X_1\cdot X_5\right) \left(X_2\cdot X_3\right)\left(X_3\cdot X_4\right)}{\left(X_2\cdot X_4\right) \left(X_1 \cdot X_3\right)\left(X_3\cdot X_5\right)}\,.
	    \label{fifthucrossratio}
	\end{equation}
	In comparison with a potentially more natural parameterization using five independent 4-point cross ratios, as in e.g.~\cite{Parikh:2019dvm,Vieira:2020xfx}, this parameterization of cross ratio space has the advantage of presenting all of our differential operators with polynomial coefficients in the $u_i$.
	
	Using the scalar representation for generators in the embedding space~\eqref{generatorsscalar}, the operators~\eqref{fivepointsQuadCasimir12}--\eqref{fivepointsvertexop} can be easily expressed as objects $D_{(X_i)}$ acting on the coordinates $X_i^A$. To obtain their action on the space of cross ratios $\mathcal{D}_{(u_i)}$, one simply conjugates the $D_{(X_i)}$ by the prefactor as follows
	\begin{equation}
		\mathcal{D}_{(u_i)}F(u_1,\dots,u_5)=\frac{1}{\Omega^{(\Delta_i)}_5}D_{\left(X_i\right)}\left(\Omega^{(\Delta_i)}_5F(u_1,\dots,u_5)\right)\,.
		\label{reductiontocrossratios5pts}
	\end{equation}
	In practical terms, the RHS above is expressed in terms of the generators~\eqref{generatorsscalar} and the expressions~\eqref{Omegafivepts}--\eqref{fifthucrossratio} of $\Omega_5^{(\Delta_i)}$ and the $u_i$'s in terms of scalar products. The LHS is then obtained by solving~(\ref{firstfourucrossratios}--\ref{fifthucrossratio}) for five different scalar products and substituting them in the RHS after the conjugation has been done; the remaining scalar products will drop out and the final answer for the LHS will be expressed only in terms of the cross ratios.

	To implement this procedure more concretely, we attach to this publication a Mathematica notebook\footnote{part of this Mathematica code is based on the one written by J. Penedones for the 2013 edition of the Mathematica Summer School in Theoretical Physics: \url{http://msstp.org/?q=node/285}.}, where we present the explicit computation of the quadratic Casimirs, as well as the final expressions one gets for the fourth-order Casimirs and the vertex operator~\eqref{fivepointsvertexop} by trivially extending the same algorithm.
	
	\medskip
	As an attempt to simplify the analytic expressions for the differential equations, it is natural to try to extend the 4-point change of coordinates of Dolan and Osborn~\cite{Dolan:2003hv,Dolan:2011dv}
	\begin{equation}
	    u=z \bar{z}\,, \qquad v=(1-z)(1-\bar{z})
	\end{equation}
    to the 5-point case. A very good candidate for this purpose is the change of coordinates
    \begin{equation}
        \begin{gathered}
        u_1=z_1 \bar{z_1}\,,\qquad u_2=(1-z_1)(1-\bar{z}_1)\,,\\
    u_3=z_2 \bar{z}_2\,, \qquad u_4=(1-z_2)(1-\bar{z}_2)\,,\\
    u_5=w(z_1-\bar{z}_1)(z_2-\bar{z}_2)+(1-z_1-z_2)(1-\bar{z}_1-\bar{z}_2)\,,
        \end{gathered}
    \end{equation}
    which leads to the simplest expressions for the quadratic Casimirs that we could find. Indeed, by introducing the notation
    \begin{gather}
    \epsilon=\frac{d-2}{2}\,,\\
        a=\frac{\Delta_1-\Delta_2}{2}\,, \qquad \tilde{a}=\frac{\Delta_5-\Delta_4}{2}\,, \qquad b=-\frac{\Delta_3}{2}\,,\\
        U_i^{(k)}=z_i^k\partial_{z_i}-\bar{z}_i^k\partial_{\bar{z}_i}\\
        V_{i,j}=\frac{z_i\bar{z}_i}{z_i-\bar{z}_i}\left( U_i^{(0)}-U_i^{(1)}+\frac{1}{z_i-\bar{z}_i}\left(1+w(z_i+\bar{z}_i-2)+\frac{z_i\bar{z}_j-z_j\bar{z}_i}{z_j-\bar{z}_j}\right)\partial_w\right)\\
        W_{i,j}=\frac{z_j+\bar{z}_j}{z_j-\bar{z}_j}U_i^{(2)}+\frac{2z_i\bar{z}_i}{z_i-\bar{z}_i}U_j^{(1)}+\frac{2z_i\bar{z}_i}{z_i-\bar{z}_i}\left(\frac{1}{z_i-\bar{z}_i}-w\frac{z_j+\bar{z}_j}{z_j-\bar{z}_j}+\frac{z_i\bar{z}_j-z_j\bar{z}_i}{(z_i-\bar{z}_i)(z_j-\bar{z}_j)}\right)\partial_w
    \end{gather}
    and the expression for the $d=1$ quadratic Casimirs
    \begin{equation}
        D_{z_1,z_2}^{(a,b)}=z_1^2(1-z_1)\partial_{z_1}^2-(a+b+1)z_1^2\partial_{z_1}-a b z_1-z_1^2 z_2 \partial_{z_1}\partial_{z_2}-a z_1 z_2 \partial_{z_2}
    \end{equation}
    one obtains the following compact expressions for the quadratic Casimirs in arbitrary dimension\footnote{these expressions differ with what one would get from~\eqref{fivepointsQuadCasimir12} and~\eqref{fivepointsQuadCasimir45} by an overall factor of $(-2).$}
    \begin{gather}
        \mathcal{D}^2_{(12)}\!=\!D_{z_1,z_2}^{(a,b)}\!+\!D_{\bar{z}_1\bar{z}_2}^{(a,b)}\!+\!2\epsilon V_{1,2}+w(1-w)\frac{(z_2-\bar{z}_2)^{1+a}}{(z_1-\bar{z}_1)^a}W_{1,2}\frac{(z_1-\bar{z}_1)^a}{(z_2-\bar{z}_2)^{1+a}}\partial_w+\frac{w}{(z_1\bar{z}_1)^a}U_1^{(2)}(z_1\bar{z}_1)^a U_2^{(1)}\,,\\
        \mathcal{D}^2_{(45)}\!=\!D_{z_2,z_1}^{(\tilde{a},b)}\!+\!D_{\bar{z}_2\bar{z}_1}^{(\tilde{a},b)}\!+\!2\epsilon V_{2,1}+w(1-w)\frac{(z_1-\bar{z}_1)^{1+\tilde{a}}}{(z_2-\bar{z}_2)^{\tilde{a}}}W_{2,1}\frac{(z_2-\bar{z}_2)^{\tilde{a}}}{(z_1-\bar{z}_1)^{1+\tilde{a}}}\partial_w+\frac{w}{(z_2\bar{z}_2)^{\tilde{a}}}U_2^{(2)}(z_2\bar{z}_2)^{\tilde{a}} U_1^{(1)}\,.
    \end{gather}
    We have also attempted similar types of factorizations for the quartic Casimirs and the vertex operator, in the spirit of the decomposition in equation $(4.14)$ of~\cite{Dolan:2011dv}; so far to no great avail.

\section{Conclusions and Outlook} 

In this paper we have initiated the construction of multipoint conformal blocks for correlation 
functions of $N$ scalar fields. More concretely, we constructed a number of independent commuting 
differential operators on $N$-point functions that matches the number of conformally invariant 
cross ratios. In consequence, these operators uniquely determine conformal blocks as bases of joint eigenfunctions into which one can expand any correlation function of $N$ scalar fields. The set of commuting differential operators depends on the choice 
of an OPE channel. We constructed them for all channel topologies, extending the results we 
announced in \cite{Buric:2020dyz}. Our construction relies on the identification of these operators with the 
Hamiltonians of certain OPE limits of integrable Gaudin models, see Section~\ref{sec:GaudinOPE}. Let us note 
that, in the special case of $d=2$, the relation between conformal blocks and Gaudin models 
is consistent with recent observations in \cite{Roehrig:2020kck} concerning ambitwistor 
realizations of scattering equations in $AdS_3$, given the close connection between conformal 
blocks and scattering amplitudes in anti-de Sitter space \cite{Hijano:2015zsa}. But it is 
important to stress that the $N$-site Gaudin Hamiltonians simplify significantly in the 
OPE limits relevant to the theory of conformal blocks. While the 4-site 
Gaudin system for $d=1$ is related to the Heun equation, for example, one obtains 
hypergeometric differential equations in the OPE limits.  

In any given channel, the set of differential operators decomposes into two subsets formed by 
Casimir differential operators on the one hand, and the novel vertex 
differential operators on the other hand. Individual vertices can give rise to multivariable integrable systems with up to $n_v(d)$ independent commuting Hamiltonians in the most generic case. All such Hamiltonians are of order 
higher than two and they have not received much attention in the physics literature so far. 
In order to develop a theory of these vertex systems it seems reasonable to start with the single-variable cases. According to formula \eqref{eq:novertrest}, all vertices that 
appear in the comb channel of an $N$-point function in $d=3,4$ contribute one single degree of freedom. The same is true for the nontrivial vertices in the comb channel of $N=5,6$-point 
functions in any $d > 2$. Therefore, single-variable vertex systems already cover many 
cases of interest. The associated vertex differential operator has already been 
constructed in \cite{Buric:2020dyz}, though not for the most general case. In the 
sequel to this work we will revisit this operator, construct it for all 
single-variable vertex systems, and show that it can be mapped to the fourth order 
Hamiltonian of an elliptic Calogero-Moser-Sutherland model that was first 
discovered about 10 years ago by Etingof, Felder, Ma and Veselov in 
\cite{etingof2021elliptic}.

There are two important directions that we will address in order to develop the theory of 
multi-point blocks. The first one concerns the extension of what we described in the previous 
paragraph to vertices with more variables, such as the central vertex in the snowflake 
channel of a $6$-point function in $d > 2$. We will discuss this in a forthcoming paper. 
The second important direction concerns the study of (joint) eigenfunctions for the vertex 
differential operator(s). Unlike the single variable of a 4-point function in $d=1$, for which 
the Casimir 
differential equation is the ordinary hypergeometric equation, even the single variable 
case of the vertex system seems to be uncharted, let alone its multi-variable extensions. 
We plan to address this challenge in the future before we combine the knowledge about 
vertex systems back into a full theory of multi-point conformal blocks. In this context, 
it would also be interesting to examine the relations of our approach with the weight shifting 
technology of multi-point blocks with non-scalar exchange recently exhibited in \cite{Poland:2021xjs}. 

While executing the entire program still requires a significant amount of work, applications 
can be pursued in parallel. This applies in particular to the multi-point bootstrap that was also advocated recently in \cite{Vieira:2020xfx}. One can expect the study of 
multi-point correlation functions to give access to new dynamical information that is not 
visible in scalar 4-point functions. The analytic bootstrap study of 4-point functions has 
provided powerful access to anomalous dimensions of double twist operators, as well as OPE 
coefficients for the operator product of two fundamental scalars into double twist operators \cite{Alday:2015eya,Alday:2016njk}. Similarly, crossing symmetry for $N=6$-point functions 
constrains the anomalous dimensions of triple twist operators and various OPEs of the 
fundamental field with double twist operators, depending on channel topologies that 
are analyzed. 
\bigskip

\noindent 
\textbf{Acknowledgements:} We are grateful to Gleb Arutyunov, Luke Corcoran, Pavel Etingof, Aleix Gimenez-Grau, 
Mikhail Isachenkov, Apratim Kaviraj, Madalena Lemos, Pedro Liendo, Junchen Rong, Joerg Teschner 
and Benoît Vicedo for useful discussions. This project received funding from the German Research 
Foundation DFG under Germany’s Excellence Strategy -- EXC 2121 Quantum Universe -- 390833306 and 
from the European Union’s Horizon 2020 research and innovation programme under the MSC grant 
agreement No.764850 “SAGEX”.

\appendix

\section{Proof of the induction in the limits of Gaudin models}
\label{app:ProofRec}

In this appendix, we detail the induction in the proof of the limit procedure of Section~\ref{sec:GaudinOPE}. We thus refer to this section for notation and definitions. Our main goal is to show that for every vertex $\rho\in V$, the Lax matrix of the $N$-sites Gaudin model satisfies the limit \eqref{eq:LimitL}. For the purposes of this appendix, it will be useful to rewrite this limit as
\begin{equation}\label{eq:LimitLapp}
\varpi^{n_\rho} \Lc_\alpha \bigl( h_\rho(z,\varpi) \bigr) \xrightarrow{\varpi \to 0} \Lc_\alpha^\rho(z) = \frac{\mathcal{T}_\alpha^{(I_{\rho,1})}}{z} + \frac{\mathcal{T}_\alpha^{(I_{\rho,2})}}{z-1},
\end{equation}
where we introduced
\begin{equation}
h_\rho(z,\varpi) = \varpi^{n_\rho}z + g_\rho(\varpi).
\end{equation}
In the left-hand side of \eqref{eq:LimitLapp}, and in all this appendix, we fix the sites $w_i$ of the Gaudin model to their value $w_i=f_i(\varpi)$ prescribed by the limit procedure of Subsection~\ref{sec:GaudinLim}. Recall that $\rho$ is either the reference vertex $\rho_\ast$, in $V'$ or in $V''$. We will treat these three cases separately.

\subsection{Reference vertex}

Let us first consider the reference vertex $\rho_\ast$. The definition of $n_{\rho_\ast}$ and $g_{\rho_\ast}$ made in Subsection~\ref{sec:GaudinLim} can be simply rewritten as
\begin{equation}
n_{\rho_\ast} = 0 \qquad \text{ and } \qquad h_{\rho_\ast}(z,\varpi) = z.
\end{equation}
We then have
\begin{equation}\label{eq:rho0}
\varpi^{n_{\rho_\ast}} \Lc_\alpha \bigl( h_{\rho_\ast}(z,\varpi) \bigr) = \Lc_\alpha(z) = \sum_{i=1}^N \frac{\mathcal{T}_\alpha^{(i)}}{z-w_i}.
\end{equation}
To proceed further, we decompose this sum over external edges $i\in \lbrace 1,\cdots,N\rbrace$ into three parts.

\paragraph{Contribution of the reference edge.} The first part is the contribution from the reference edge $N$, with corresponding site $w_N=f_N(\varpi)=\varpi^{-1}$, according to eq. \eqref{eq:wHighest}. It simply reads
\begin{equation}\label{eq:ContribN}
\frac{\mathcal{T}_\alpha^{(N)}}{z-w_N} = \frac{\varpi \,\mathcal{T}_\alpha^{(N)}}{\varpi \, z - 1} \xrightarrow{\varpi\to 0} 0
\end{equation}
and thus does not contribute to the limit of \eqref{eq:rho0} when $\varpi\to 0$.

\paragraph{Contribution of the subtree $T'$.} The second part is the contribution from the subtree $T'$ attached to $e'$, for which we should distinguish the cases where $e'$ is external or not. If $e'$ is external (in which case $T'$ is trivial and $e'$ is the only contribution from this subtree), recall from the first case of eq. \eqref{eq:w'} that we then have $w_{e'}=f_{e'}(\varpi)=\varpi$, so that the contribution to \eqref{eq:rho0} simply is
\begin{equation}\label{eq:Contribe'}
\frac{\mathcal{T}_\alpha^{(e')}}{z-\varpi} \xrightarrow{\varpi\to 0} \frac{\mathcal{T}_\alpha^{(e')}}{z}.
\end{equation}
Note that as $e'$ is external, we have $E'=\lbrace e' \rbrace$ so that the generators $\mathcal{T}_\alpha^{(e')}$ coincide with $\mathcal{T}_\alpha^{(E')}$.

If $e'$ is intermediate in the initial diagram then the contribution from the subtree $T'$ comes from the edges $i\in E' \subset \underline{N}$, with corresponding sites $w_{i} = f_i(\varpi) = \varpi\,f'_{i}(\varpi)$ as in the second line of eq. \eqref{eq:w'}. By construction, the functions $f_i'(\varpi)$ associated with the external edges $i\in E'$ stay finite when $\varpi\to0$: indeed, as we supposed $T'$ non-trivial, the edges $i\in E'$ are not the reference edge $e'$ of $T'$ and are thus associated with polynomial functions $f'_i(\varpi)$. Thus, we get that the corresponding sites $w_{i} = \varpi\,f'_{i}(\varpi)$ in the initial tree tend to 0 when $\varpi\to 0$. In this limit, the contribution of this subtree $T'$ to the sum \eqref{eq:rho0} then simply becomes
\begin{equation}
\sum_{i \in E'} \frac{\mathcal{T}_\alpha^{(i)}}{z-w_{i}} \xrightarrow{\varpi\to 0} \frac{1}{z} \sum_{i \in E'} \mathcal{T}_\alpha^{(i)}.
\end{equation}
To conclude, let us observe that the sum in the right-hand side of the above equation is by definition $\mathcal{T}_\alpha^{(E')}$. Combined with the eq. \eqref{eq:Contribe'} and the discussion that followed it for the case where $e'$ is external, we thus observe that in both cases $e'$ external and internal, the contribution of the subtree $T'$ in the limit $\varpi\to0$ of eq. \eqref{eq:rho0} is simply
\begin{equation}\label{eq:ContribE'}
\sum_{i \in E'} \frac{\mathcal{T}_\alpha^{(i)}}{z-w_{i}} \xrightarrow{\varpi\to 0} \frac{\mathcal{T}_\alpha^{(E')}}{z}.
\end{equation}

\paragraph{Contribution of the subtree $T''$.} A similar argument applies to the contribution of the right subtree $T''$, attached to $e''$. If the latter is external, this contribution is simply equal to
\begin{equation}
\frac{\mathcal{T}_\alpha^{(e'')}}{z-1},
\end{equation}
since we then have $w_{e''}=1$ -- see the first line of eq. \eqref{eq:w''}. If $e''$ is intermediate, the corresponding sites $w_{i}$, $i \in E''$, are given by the second line of eq. \eqref{eq:w''} and read $w_{i} = f_i(\varpi) = 1 + \varpi\,f''_{i}(\varpi)$, which thus tend to 1 when $\varpi \to 0$. In both cases, we find that the contribution to \eqref{eq:rho0} is given by
\begin{equation}\label{eq:ContribE''}
\sum_{i \in E''} \frac{\mathcal{T}_\alpha^{(i)}}{z-w_{i}} \xrightarrow{\varpi\to 0} \frac{\mathcal{T}_\alpha^{(E'')}}{z-1},
\end{equation} 
where in the first case $\mathcal{T}_\alpha^{(E'')}=\mathcal{T}_\alpha^{(e'')}$ while in the second case $\mathcal{T}_\alpha^{(E'')}$ is a composite operator formed by the sum of $\mathcal{T}_\alpha^{(i)}$ for $i\in E''$.

\paragraph{Conclusion.} Summing the contributions \eqref{eq:ContribN}, \eqref{eq:ContribE'} and \eqref{eq:ContribE''} in the limit of eq. \eqref{eq:rho0} when $\varpi\to0$, we then find
\begin{equation}
\varpi^{n_{\rho_\ast}} \Lc_\alpha \bigl( h_{\rho_\ast}(z,\varpi) \bigr) \xrightarrow{\varpi\to0} \frac{\mathcal{T}_\alpha^{(E')}}{z} + \frac{\mathcal{T}_\alpha^{(E'')}}{z-1}.
\end{equation}
To conclude, we observe that the labeling of branches at vertices made in Subsection~\ref{sec:GaudinLim} using the plane rooted tree representation $T$ of the diagram implies that $E'=I_{\rho_\ast,1}$ and $E''=I_{\rho_\ast,2}$. This shows that the limit \eqref{eq:LimitLapp} is satisfied for the reference vertex $\rho_\ast$.

\subsection{Vertices in \texorpdfstring{$\boldmath{V'}$}{V'}.}

Let us now consider the case of a vertex $\rho \in V'$ in the subtree attached to $e'$. Note that the existence of this vertex requires the subtree $T'$ to be non-trivial and thus the edge $e'$ to be intermediate in the initial diagram. The definition of $n_\rho$ and the recursion relation for $g_\rho(\varpi)$ in the first line of eq. \eqref{eq:gRec} can be rewritten as
\begin{equation}
n_\rho = n'_\rho + 1 \qquad \text{ and } \qquad h_\rho(z,\varpi) = \varpi\,h'_\rho(z,\varpi),
\end{equation}
where we have defined $h'_\rho(z,\varpi)=\varpi^{n'_\rho}z + g'_\rho(\varpi)$ as the equivalent of $h_\rho(z,\varpi)$ for the subtree $T'$. We thus have
\begin{equation}\label{Eq:Lrho'}
\varpi^{n_{\rho}} \Lc_\alpha \bigl( h_\rho(z,\varpi) \bigr) = \varpi^{n'_\rho+1} \Lc_\alpha \bigl( \varpi\,h'_\rho(z,\varpi) \bigr) = \sum_{i=1}^N \frac{\varpi^{n'_\rho+1}\mathcal{T}_\alpha^{(i)}}{\varpi\,h'_\rho(z,\varpi)-w_i}.
\end{equation}
We will once again separate this expression in three parts, coming from the contributions of the reference edge $N$ and the two subtrees.

\paragraph{Contribution of the reference edge.} Let us start with the edge $N$, with associated site $w_N=f_N(\varpi)=\varpi^{-1}$. Its contribution to \eqref{Eq:Lrho'} in the limit $\varpi\to0$ is then given by
\begin{equation}
\frac{\varpi^{n'_\rho+1}\mathcal{T}_\alpha^{(N)}}{\varpi\,h'_\rho(z,\varpi) - w_N} = \frac{\varpi^{n'_\rho+2}\mathcal{T}_\alpha^{(N)}}{\varpi^2\,h'_\rho(z,\varpi) - 1} \; \xrightarrow{\varpi\to 0} \, 0.
\end{equation}

\paragraph{Contribution of the subtree $T''$.} Let us now consider the contribution coming from the subtree $T''$ attached to $e''$. If $e''$ is external in the initial tree $T$, then we simply have $w_{e''} = 1$ according to the first line of eq. \eqref{eq:w''} and the contribution to \eqref{Eq:Lrho'} is
\begin{equation}
\frac{\varpi^{n'_\rho+1}T_\alpha^{(e'')}}{\varpi\,h'_\rho(z,\varpi) - 1} \; \xrightarrow{\varpi\to 0} \, 0.
\end{equation}
If $e''$ is initially intermediate, then the contribution comes from the external edges $i \in E''\subset\underline{N}$ whose corresponding sites are given by the second line of eq. \eqref{eq:w''} to be $w_{i} = f_i(\varpi) = 1 + \varpi\,f''_{i}(\varpi)$. In particular, these tend to 1 when $\varpi\to0$. Thus, in this case, the contribution of the subtree $T''$ to \eqref{Eq:Lrho'} also vanishes:
\begin{equation}
\sum_{i \in E''} \frac{\varpi^{n'_{\rho}+1}\mathcal{T}_\alpha^{(i)}}{\varpi\,h'_\rho(z,\varpi) - w_{i}} \; \xrightarrow{\varpi\to 0} \, 0.
\end{equation}

\paragraph{Contribution of the subtree $T'$.} Thus, the only non-vanishing contribution to \eqref{Eq:Lrho'} in the limit $\varpi\to0$ comes from the subtree $T'$ attached to $e'$. Recall that $T'$ is non-trivial since it possesses a vertex $\rho$: the external edges $i\in E'\subset\underline{N}$ are thus different from $e'$ and are associated with the sites $w_{i} = f_i(\varpi) = \varpi f'_{i}(\varpi)$, according to the second line of eq. \eqref{eq:w'}. The contribution of $T'$ to \eqref{Eq:Lrho'} then reads
\begin{equation}\label{Eq:Contrib'}
\sum_{i \in E'} \frac{\varpi^{n'_\rho+1}\mathcal{T}_\alpha^{(i)}}{\varpi\,h'_\rho(z,\varpi) - w_{i}} = \sum_{i \in E'} \frac{\varpi^{n'_\rho}\mathcal{T}_\alpha^{(i)}}{h'_\rho(z,\varpi) - f'_{i}(\varpi)}.
\end{equation}
The Gaudin Lax matrix of the subtree $T'$ (the one of the full tree, not the ones associated with vertices) is given by eq. \eqref{eq:L'} and thus can be rewritten as
\begin{equation}
\Lc'_\alpha(z) = \sum_{i\in E'} \frac{\mathcal{T}_\alpha^{(i)}}{z-f'_{i}(\varpi)} + \frac{\varpi\,\mathcal{T}_\alpha^{(e')}}{\varpi\,z-1},
\end{equation}
where we used the fact that $f'_{e'}(\varpi)=\varpi^{-1}$ since $e'$ is the reference edge of $T'$. Thus, we can rewrite the above contribution \eqref{Eq:Contrib'} as
\begin{equation}
\sum_{i \in E'} \frac{\varpi^{n'_\rho+1}\mathcal{T}_\alpha^{(i)}}{\varpi\,h'_\rho(z,\varpi) - w_{i}} = \varpi^{n'_\rho} \Lc'_\alpha \bigl( h'_{\rho}(z,\varpi) \bigr) - \frac{\varpi^{n'_\rho+1}\mathcal{T}_\alpha^{(e')}}{\varpi \,h'_\rho(z,\varpi) - 1}.
\end{equation}
The second term vanishes in the limit $\varpi \to 0$. Moreover, by the induction hypothesis \eqref{eq:LimitL'} for the subtree $T'$, the first term tends to $\Lc_\alpha^\rho(z)$ in this limit. In conclusion, we thus get
\begin{equation}\label{eq:contrib'lim}
\sum_{i \in E'} \frac{\varpi^{n'_\rho+1}\mathcal{T}_\alpha^{(i)}}{\varpi\,h'_\rho(z,\varpi) - w_{i}} \xrightarrow{\varpi\to 0}  \Lc_\alpha^\rho(z).
\end{equation}

\paragraph{Conclusion.} Since the contribution \eqref{eq:contrib'lim} is the only non-vanishing term in the limit of eq. \eqref{Eq:Lrho'} when $\varpi\to 0$, we thus get in the end that
\begin{equation}
\varpi^{n_\rho} \Lc_\alpha \bigl( h_\rho(z,\varpi) \bigr) \xrightarrow{\varpi\to 0}  \Lc_\alpha^\rho(z),
\end{equation}
as required. This indeed shows that the limit \eqref{eq:LimitLapp} is satisfied for vertices $\rho\in V'$, using the induction hypothesis \eqref{eq:LimitL'} that a similar property also holds in the subtree $T'$.

\subsection{Vertices in \texorpdfstring{$\boldmath{V''}$}{V''}}

Let us finally consider a vertex $\rho\in V''$. The procedure here will resemble the one in the previous subsection so we will not describe it in detail. The definition of $n_\rho$ and the recursion relation for $g_\rho(\varpi)$ in the second line of eq. \eqref{eq:gRec} can be rewritten as
\begin{equation}
n_\rho = n''_\rho + 1 \qquad \text{ and } \qquad h_\rho(z,\varpi) = 1 + \varpi\,h''_\rho(z,\varpi).
\end{equation}
The main difference with the case of a vertex in $V'$ is that we introduced a shift by $1$ in the expression of $h_\rho(z,\varpi)$. The effect of this shift is that in the computation of the limit of $\varpi^{n_\rho} \Lc_\alpha \bigl( h_\rho(z,\varpi) \bigr)$ when $\varpi\to 0$, the only non-vanishing contribution now comes from the subtree $T''$ and not from $T'$. More precisely, using the fact that the shift by $1$ cancels with the $1$ in the recursive expression \eqref{eq:w''} of the sites $w_{i}$, $i\in E''$, we find in the end that
\begin{equation}
\lim_{\varpi\to 0} \varpi^{n_\rho} \Lc_\alpha \bigl( h_\rho(z,\varpi)\bigr) = \lim_{\varpi\to 0} \varpi^{n''_\rho} \Lc''_\alpha \bigl( h''_\rho(z,\varpi)\bigr) = \Lc_\alpha^\rho(z),
\end{equation}
with $\Lc''_\alpha$ the Lax matrix \eqref{eq:L''} associated with the subtree $T''$, and where the last equality follows from the induction hypothesis \eqref{eq:LimitL''} for the subtree $T''$. This then completes the proof of the induction.

\section{Embedding space formalism and index-free notation}
\label{appendix:Embeddingspace}
In this section we will briefly review the embedding space formalism of~\cite{Costa:2011mg}, as well as a slight generalization of their index-free notation that we are going to use in the proof of Appendix~\ref{appendix:prooflemma}. We will in particular focus on scalar and tensorial representations. In short, the embedding space formalism makes use of the isomorphism between the conformal group in $d$ dimensions and the Lorentz group in $d+2$ dimensions -- both corresponding to $SO(d+1,1)$ -- to map objects from one space to the other. Points $x\in\mathbb{R}^d$ are mapped to $(d+2)$-dimensional vectors $X^A=(X^+,X^-,X^\mu)$, here presented in light-cone coordinates with metric $ds^2 = -dX^- dX^+ + \delta_{\mu\nu} dX^\mu dX^\nu$, and to make the degrees of freedom match, these vectors are constrained to the projective light-cone
\begin{equation}
	X^A=X^+(1,x^2,x^\mu)\simeq(1,x^2,x^\mu)\,.
\end{equation} 
With this mapping, scalar fields become homogeneous objects of degree $-\Delta$ with respect to the coordinates $X^A$ they depend on
	\begin{equation}
	    \mathcal{O}_\Delta(\lambda X)= \lambda^{-\Delta} \mathcal{O}_\Delta(X)\,.
	\end{equation}
	
	This has the big advantage that the conformal generators, some of which are non-linear, once mapped to the embedding space, acquire all the same linear form, namely the one of Lorentz generators:
	\begin{equation}
		\mathcal{T}_{[AB]}=X_{A}\partial_B-X_{B}\partial_A\,,
		\label{generatorsscalar}
	\end{equation}
	where $X_A$ is obtained by lowering the index of $X^A$ by contraction with the flat metric $\eta_{AB}$.
	This not only simplifies the analytic expressions that one works with, but also allows one to carry out computer algebra computations in a much faster and general way through the implementation of indices; this proves very useful for the computation of operators for scalar multipoint blocks like those of Section~\ref{section:fivepoints}. 
	
	When one considers fields in spinning representations, labeled by $L$ spins $l_1\ge l_2\ge \dots \ge l_L$
	\begin{equation}
	    \mathcal{O}_{\Delta,\{l_i\}}\equiv \mathcal{O}_{\left(A^{(1)}_1 \dots A^{(1)}_{l_1}\right)\dots \left(A^{(L)}_1\dots A^{(L)}_{l_L}\right)}\,,
	    \label{eq:mixedsymmetrysymmtrz}
	\end{equation}
	where indices in the same bracket are symmetrized over, one also requires them to be transverse with respect to any index
	\begin{equation}
    X^{A_i^{(j)}}\mathcal{O}_{\left(A^{(1)}_1 \dots A^{(1)}_{l_1}\right)\dots A_i^{(j)}\dots \left(A^{(L)}_1\dots A^{(L)}_{l_L}\right)}=0\,,
\end{equation}
    and traceless with respect to any pair of indices
    \begin{equation}
    \eta^{A^{(j_1)}_{i_1} A^{(j_2)}_{i_2}}\mathcal{O}_{A^{(1)}_1 \dots A^{(j_1)}_{i_1} \dots A^{(j_2)}_{i_2}\dots A^{(L)}_{l_L}}(X)=0\,.
\end{equation}
    Expressions involving these tensorial representations can be further simplified by the use of an index-free notation in embedding space. One can implement this by either making the antisymmetries of the tensors manifest with fermionic variables, as was first done in~\cite{Costa:2014rya}, or by making the symmetries manifest with bosonic variables, in a form like Appendix~B of~\cite{Kologlu:2019mfz}. In agreement with our choice of representation for~\eqref{eq:mixedsymmetrysymmtrz}, we will keep the symmetries manifest and use the latter formalism. The core idea is to remove indices from the tensorial expressions~\eqref{eq:mixedsymmetrysymmtrz} by converting them to polynomials on $m$ auxiliary vectors $Z_i\in \mathbb{C}^{d+2}$ through the contractions
    \begin{equation}
        \mathcal{O}_{\Delta,\{l_i\}}(X,\{Z_i\})\equiv\mathcal{O}_{\left(A^{(1)}_1 \dots A^{(1)}_{l_1}\right)\dots \left(A^{(L)}_1\dots A^{(L)}_{l_L}\right)}\left(Z_1^{A_1^{(1)}}\cdots Z_1^{A_{l_1}^{(1)}}\right)\cdots\left(Z_L^{A_1^{(L)}}\cdots Z_L^{A_{l_L}^{(L)}}\right)\,.
    \end{equation}
	To keep everything consistent with the conformal covariance, tracelessness and transversality of the tensors, the coordinate vectors need to satisfy
	\begin{equation}
	    X^2=X\cdot Z_i=Z_i\cdot Z_j=0\,, \qquad \qquad \forall i,j\in\{1,2,\dots,L\}\,,
	\end{equation}
	while the fields need to have polynomial dependence on all the $Z_i$, be homogeneous in every coordinate
	\begin{equation}
    \mathcal{O}_{\Delta,\{l_i\}}\!\left(\lambda_0 X, \{ \lambda_i Z_i\}\right)=\lambda_0^{-\Delta}\lambda_1^{l_1}\cdots \lambda_L^{l_L} \mathcal{O}_{\Delta,\{l_i\}}(X,\{Z_i\})\,,
    \label{tensorshomogeneityconditions}
\end{equation}
    and invariant under the following gauge transformations
    \begin{equation}
    \mathcal{O}_{\Delta,\{l_i\}}\!\Bigl(X,\Bigl\{Z_i+\beta_{i,0} X +\sum_{j<i} \beta_{i,j} Z_j\Bigr\}\Bigr)=\mathcal{O}_{\Delta,\{l_i\}}(X,\{Z_i\})\,, \qquad \forall \beta_{ij}\in \mathbb{C}\,.
    \label{tensorsgaugeinvariance}
\end{equation}
    Note how the $Z_i$ are not all on equal footing, but there is rather a hierarchy of gauge transformations~\eqref{tensorsgaugeinvariance} that respects the hierarchy of spins $l_1\ge \dots \ge l_L$: the $Z_i$ associated to larger spins are less constrained than the ones for smaller spins. In fact, a function $\mathcal{O}_{\Delta,\{l_i\}}(X,\{Z_i\})$ satisfying the constraints~\eqref{tensorsgaugeinvariance} is a section of a line bundle over the flag manifold $\mathrm{SO}(1,d+1)/P_{d,L}$\footnote{this is a ``partial'' flag manifold if $L+1< r_d$, and a ``full'' flag manifold if $L+1 = r_d$.}, $P_{d,L} = ( \mathrm{SO}(1,1) \times (\mathrm{SO}(2) \times ( ... \times ( \mathrm{SO}(2) \times \mathrm{SO}(d-2L) ) \ltimes \mathbb{C}^{d-2L} ) \ltimes \mathbb{C}^{d-2L+2} ) ... ) \ltimes \mathbb{C}^{d-2} ) \ltimes \mathbb{R}^d)$, each point of which is labeled by an embedding of vector spaces $\{0\} \subset \mathrm{Span}(X) \subset \mathrm{Span}(X,Z_1)\subset ... \subset \mathrm{Span}(X,Z_1,...,Z_L) \subset \mathbb{C}^{d+2}$. 
    
	The action of the conformal generators on these fields is also modified to take into account the $Z_i$ dependence, so~\eqref{generatorsscalar} becomes
	\begin{equation}
    \mathcal{T}_{[AB]}=X_A\pdv{X^B}+\sum_{i=1}^L Z_{i\,A}\pdv{Z^{B}_i} - (A\leftrightarrow B)\,.
    \label{generatorsspinningembedding}
    \end{equation}
	Using this type of representation for generators and fields, it becomes much simpler to obtain Casimir and vertex differential operators for spinning external fields.
	
	\subsection{Classical Embedding space}
	\label{sect:ClassicalEmbSp}
	For the proof in Appendix~\ref{appendix:prooflemma} we are going to need a classical version of the embedding space formalism for the spinning fields that we have just discussed. For this purpose, we introduce $P_A$ as the conjugate momentum to $X^A$, and $Q_A^i$ as conjugate momenta of the auxiliary variables $Z_i^A$.
	
	To see which constraints are imposed on these classical variables, one can check the action of the operators
	\begin{equation}
	    X\cdot \pdv{X}\,, \qquad X\cdot \pdv{Z_i}\,, \qquad Z_i\cdot \pdv{X}\,, \qquad Z_i\cdot \pdv{Z_j}
	    \label{operatorsscalarproducts}
	\end{equation}
	on fields or correlation functions to get conditions that need to be satisfied by scalar products of coordinates and momenta. Some of the operators in~\eqref{operatorsscalarproducts} are in fact evaluated to constants on correlation functions. By imposing the same behavior when replacing derivatives with momenta, one obtains the following relations for phase space variables:
\begin{equation}
\begin{gathered}
    X^2=Z_i^2=Z_i\cdot Z_j=0\,,\\
    X\cdot Z_i=X\cdot Q_i=0\,,\\
    Z_i\cdot Q_j=0 \quad\forall\, i<j\,,\\
    X\cdot P=-\Delta\,, \quad Z_i\cdot Q_i=l_i\,.
    \end{gathered}
    \label{conditionsclassicalembeddingsp}
\end{equation}
	One can directly verify that the conditions~\eqref{conditionsclassicalembeddingsp} combined with the classical version of the generators~\eqref{generatorsspinningembedding}
	\begin{equation}\label{eq:TABspin}
	    \bar{\mathcal{T}}_{[AB]}=X_A P_B+\sum_{i=1}^L Z_{i\,A}Q_{i\,B} - (A\leftrightarrow B)
	\end{equation}
	lead to the correct classical Casimirs associated to mixed-symmetry tensors:
	\begin{equation}\label{eq:ClassicalCas}
	    \mathcal{C}\textit{\!as\,}^{2p}=2\Delta^{2p}+2\sum_{i=1}^{L} l_i^{2p}\,.
	\end{equation}

\section{Relations among vertex differential operators} 
\label{appendix:prooflemma}

Our goal in this appendix is to justify the relations \eqref{eq:lemmaf} and \eqref{eq:lemmas}
for the total symbols of our vertex differential operators. There are many ways to derive these 
relations. Here we shall follow a more pedestrian approach that does not require much background from 
representation theory. 

In order to derive the relations \eqref{eq:lemmaf}, \eqref{eq:lemmas} we first note that these 
were formulated in terms of the coordinates and momenta of the external scalar fields. The representation of the conformal algebra that is associated to the index set $I$ decomposes 
into an infinite number of spinning representations. Each of the irreducible components can 
be prepared in embedding space formalism (see appendix \ref{appendix:Embeddingspace}). Here we shall study the relations 
\eqref{eq:lemmaf} for a given irreducible component so that the coefficients $\varrho_{f,s}$ are 
functions of the associated weight and spins rather than functions of symbols of the Casimir 
differential operators. 

After these introductory comments let us approach relation \eqref{eq:lemmaf} by considering 
the simplest example in which the intermediate irreducible representation is scalar. We can  
construct explicitly the first three matrices $( \bar{\mathcal{T}}^{n})^{A}_{\;\;\,B}$\footnote{Here 
and in the discussion below we shall drop the subscript $f$.} in the classical embedding space that 
we introduced in Appendix~\ref{sect:ClassicalEmbSp}:
\begin{equation}
\begin{gathered}
    \left( \bar{\mathcal{T}}^1\right)^{A}_{\;\;\,B} = X^A P_B-X_B P^A\\
    \left( \bar{\mathcal{T}}^2\right)^{A}_{\;\;\,B} =\Delta\left(X^A P_B +X_B P^A\right)-P^2X^AX_B\\
    \left( \bar{\mathcal{T}}^3\right)^{A}_{\;\;\,B} =\Delta^2 \left( X^A P_B- X_B P^A\right)=\Delta^2  \left( \bar{\mathcal{T}}^1\right)^{A}_{\;\;\,B} \,.
    \label{scalardependences}
\end{gathered}
\end{equation}
It is then clear that, when considering scalar representations, the powers of generators $ \bar{\mathcal{T}}^n$ with $n\ge 3$ will be dependent on lower powers of the generators. This 
directly implies that any vertex operator of the type~\eqref{eq:vdoalt} that contains a power 
of a scalar generator higher than two will become dependent on operators of lower order, e.g.
\begin{equation}
     \bar{\mathcal{T}}^3\cdot \mathcal{B}\propto  \bar{\mathcal{T}} \cdot \mathcal{B}\,.
\end{equation}
We would now like to prove that something analogous to~\eqref{scalardependences} is valid for 
representations of higher depth $\dep$ and higher powers, respectively. Let us then consider 
the generators for a mixed symmetry tensor of depth $\dep$, 
\begin{equation}
     \bar{\mathcal{T}}_{[AB]}=X_A P_B+\sum_{i=1}^{\dep-1} Z_{i\,A}Q_{i\,B} - (A\leftrightarrow B);
\end{equation}
we expect the $n$-fold contractions of generators $ \bar{\mathcal{T}}^n$ to be independent up to 
power $n=2\dep$, with the first dependent object produced at power $n=2\dep+1$. To prove this, let 
us focus on a specific matrix entry $( \bar{\mathcal{T}}^n)^{A}_{\;\;\,B} $ of the powers $\bar{\mathcal{T}}^n$ and construct the following submatrix of the Jacobian for fixed indices $A$, $B$ and $C$:
\begin{equation}
    \left(
    \begin{array}{cccccc}
    \pdv{{( \bar{\mathcal{T}}^1)}^{A}_{\;\;\,B} }{X^C}     & \pdv{{( \bar{\mathcal{T}}^1)}^{A}_{\;\;\,B} }{P^C} & \pdv{{( \bar{\mathcal{T}}^1)}^{A}_{\;\;\,B} }{Z_1^C} & \pdv{{( \bar{\mathcal{T}}^1)}^{A}_{\;\;\,B} }{Q_1^C} & \cdots & \pdv{{( \bar{\mathcal{T}}^1)}^{A}_{\;\;\,B} }{Q_{\dep-1}^C}\\
        \vdots &\vdots & \vdots &\vdots & &\vdots \\
    \pdv{{( \bar{\mathcal{T}}^{2\dep+1})}^{A}_{\;\;\,B} }{X^C}     & \pdv{{( \bar{\mathcal{T}}^{2\dep+1})}^{A}_{\;\;\,B} }{P^C} & \pdv{{( \bar{\mathcal{T}}^{2\dep+1})}^{A}_{\;\;\,B} }{Z_1^C} & \pdv{{( \bar{\mathcal{T}}^{2\dep+1})}^{A}_{\;\;\,B} }{Q_1^C} & \cdots & \pdv{{(\bar{\mathcal{T}}^{2\dep+1})}^{A}_{\;\;\,B}}{Q_{\dep-1}^C}
    \end{array}
    \right)\,.
    \label{Jacobianmatrixindependence}
\end{equation}
If what we argued above holds, we should be able to see that some $2\dep$-minors 
of~\eqref{Jacobianmatrixindependence} are equal to zero, while the same matrix with the last 
row dropped out has all nonzero $2\dep$-minors. We checked this with Mathematica symbolically 
for the case of symmetric traceless tensors, and numerically for mixed symmetry tensors of 
depth $\dep \le 6$, exhausting all tensorial representations that are allowed in the range 
of dimensions of known CFTs. One can use a similar reasoning to show eq.\ \eqref{eq:lemmas}. 

Let us finally comment on a more conceptual interpretation of the results of this appendix. We consider first the case of a scalar representation, whose generators can be gathered in the matrix $ (\bar{\mathcal{T}}_f)^{A}_{\;\;\,B} = X^A P_B-X_B P^A$ in the fundamental representation. Using the relations $X^A X_A=0$ and $X^A P_A = -\Delta$, one can show that the matrix $\bar{\mathcal{T}}_f$ is diagonalisable, with eigenvalues $\Delta$, $-\Delta$ and $0$ (with multiplicity $d$). It is a standard result of linear algebra that $\bar{\mathcal{T}}_f$ is then annihilated by the polynomial with simple roots equal to these eigenvalues, namely $T(T-\Delta)(T+\Delta)=T^3-\Delta^2T$: we recover this way that $\bar{\mathcal{T}}_f^3 = \Delta^2 \bar{\mathcal{T}}_f$.  

A similar argument can be formulated for a representation with higher depth $\dep=L+1$, characterized by a weight $\Delta$ and $L$ spins $l_1,\dots,l_L$. The (symbols of the) generators of this representation are given by eq. \eqref{eq:TABspin} and can be gathered in a matrix $\bar{\mathcal{T}}_f$ valued in the fundamental representation. The traces of odd powers of $\bar{\mathcal{T}}_f$ vanish, while the traces of even powers are given by the classical Casimirs \eqref{eq:ClassicalCas}. These traces are the Newton sums $\sum_{i=1}^{d+2} \lambda_i^p$ of the eigenvalues $\lambda_1,\cdots,\lambda_{d+2}$ of $\bar{\mathcal{T}}_f$ and thus determine these eigenvalues uniquely (up to permutation). More precisely, we find that $\bar{\mathcal{T}}_f$ has eigenvalues $\Delta,-\Delta,l_1,-l_1,\dots,l_L,-l_L$ and $0$ (with multiplicity $d-2L$). If we suppose that $\bar{\mathcal{T}}_f$ is diagonalizable, it is then annihilated by the polynomial with simple roots equal to these eigenvalues, hence
\begin{equation}\label{eq:RelT}
\bar{\mathcal{T}}_f \bigl( \bar{\mathcal{T}}_f^2 - \Delta^2 \bigr) \bigl( \bar{\mathcal{T}}_f^2 - l_1^2 \bigr)\cdots\bigl( \bar{\mathcal{T}}_f^2 - l_L^2 \bigr) = 0.
\end{equation}
This shows that the power $\bar{\mathcal{T}}_f^{2L+3}=\bar{\mathcal{T}}_f^{2\dep+1}$ is expressible in terms of lower power $\bar{\mathcal{T}}_f^n$, $n\leq 2\dep$, as expected. Let us finally note that the coefficients in the relation \eqref{eq:RelT} are elementary symmetric polynomials in the variables $(\Delta^2,l_1^2,\dots,l_L^2)$: by the Newton identities, these coefficients are then also polynomials in the Newton sums $\Delta^{2p}+\sum_{i=1}^L l_i^{2p}$ and thus in the values \eqref{eq:ClassicalCas} of the classical Casimirs. This ensures that the coefficients $\varrho^{(n,m)}_{f;AB}$ in eq. \eqref{eq:lemmaf} are polynomials in the total symbols of the Casimir operators.

\end{document}